\newif\ifSpringer
\newcommand{\springer}[2][]{\ifSpringer#2\else#1\fi}

\Springerfalse

\ifSpringer
\documentclass{llncs}
\else
\documentclass[a4paper,11pt]{article}
\fi

\usepackage{kolarto}

\ifSpringer
\usepackage[firstpage]{draftwatermark}  % free badge placement
\fi

\ifSpringer
\let\llncssubparagraph\subparagraph
\let\subparagraph\paragraph
\usepackage{titlesec}
\let\subparagraph\llncssubparagraph
\titlespacing*{\subsubsection}{0pt}{15pt plus 4pt minus 4pt}{0.5em plus 0.22em minus 0.1em}
\titlespacing*{\paragraph}{0pt}{8pt plus 4pt minus 4pt}{0.5em plus 0.22em minus 0.1em}

\AtBeginDocument{\setlength\abovedisplayskip{8pt plus 3pt minus 4pt}}
\AtBeginDocument{\setlength\belowdisplayskip{\abovedisplayskip}}

\else
\fi % Springer

\usepackage{enumitem}
\setlist[itemize]{noitemsep, topsep=4pt plus 2pt minus 2pt}

\Finaltrue
\Fulltrue
\newacr{etcs}{ETCS}{European Train Control System}

\title{Railway Scheduling Using\\
Boolean Satisfiability Modulo Simulations}

\def\InstA{Czech Technical University in Prague, Faculty of Information Technology}
\def\InstB{Institute of Computer Science of the Czech Academy of Sciences}
\def\OrcidA{0000-0002-7207-5197}
\def\OrcidB{0000-0003-1710-1513}

\author{\open{%
Tom{\'a}{\v s} Kol{\'a}rik%
\ifSpringer\inst{1}\else\footnote{\InstA. ORCID: \OrcidA.}\fi%
\springer{\orcidID{\OrcidA}}
\springer[ and]{\and}
Stefan Ratschan%
\ifSpringer\inst{2}\else\footnote{\InstB. ORCID: \OrcidB.}\fi%
\springer{\orcidID{\OrcidB}}
}}

\ifSpringer
\institute{\open{\InstA\and\InstB}}
\fi

\begin{document}
\maketitle

\ifSpringer
\SetWatermarkText{\raisebox{12.5cm}{%
    \hspace{0.1cm}%
    \href{https://doi.org/10.5281/zenodo.7351881}{\includegraphics{fm_available}}% Update DOI to reflect your artefact
    \hspace{10.5cm}%
}}
\fi

%%%%%%%%%%%%%%%%%%%%%%%%%%%%%%%%%%%%%%%%%%%%%%%%%%%%%%%%%

\blind{\vspace{-2.5em}}

\begin{abstract}
\hide{%
SAT solvers are convenient tools to be used in the area of scheduling (or planning).
Except of Boolean constraints, it is often important to~model dynamic phenomena,
which is handled by simulation tools in industry.
}

Railway scheduling is a problem that exhibits both non-trivial discrete and continuous behavior.
In this paper, we
\full{%
simulate train networks at a~low level,
where a~number of timing and ordering constraints can appear.
We
}%
model this problem using
a~combination of SAT and \acrl{ode}s (SAT modulo ODE).
In addition, we adapt \open[an~]{our }existing method
for solving such problems in such a~way that the resulting solver is competitive
with methods based on dedicated railway simulators while being more general and extensible.
\end{abstract}

%%%%%%%%%%%%%%%%%%%%%%%%%%%%%%%%%%%%%%%%%%%%%%%%%%%%%%%%%
%%%%%%%%%%%%%%%%%%%%%%%%%%%%%%%%%%%%%%%%%%%%%%%%%%%%%%%%%

\section{Introduction}\label{s:intro}
Existing benchmark problems for SAT modulo ODE~\cite{isat-ode-art,dreal-art}
do not exhibit complex discrete state space.
In this paper, we develop a~benchmark problem that combines
a~non-trivial propositional part with differential equations.
Moreover, we apply and improve a~corresponding algorithm~\cite{my-tap-art}
that tightly integrates SAT and numeric simulations of differential equations.
The resulting tool is available online~\opencite{unsot}.

\full{%
Scheduling is a~native satisfiability problem,
where a~plan that meets all criteria is sought.
It starts to be challenging when plans should also cover
complex dynamic phenomena,
usually in the form of differential equations.
Such dynamic models, typically of cyber-physical systems,
can be simulated using numerical solvers,
such as Xcos, Simulink, or SpaceEx~\cite{spaceex-art}.
However, automatic verification tasks, or planning tasks,
are difficult with such complex models,
especially if discrete state space is huge and complicated,
which is typical for SAT.
In the area of railway transport,
precise scheduling can become to~play an~important role
within upcoming autonomous traffic control systems.
}

\full[The]{Our} benchmark problem comes from the domain of railway scheduling,
and is inspired by an approach to railway design capacity analysis~\cite{rail-art},
that combines a~SAT solver with a~railway simulator.
The authors of that approach, referring to SAT modulo non-linear real arithmetic,
``found these solvers insufficiently scalable for real-world problem sizes.''
Our experiments show that it indeed \emph{is} possible to realistically handle continuous dynamics
in the railway domain directly by SAT modulo theory solvers.
A~major difficulty lies in modeling the fact that
trains sometimes have to switch to a~deceleration phase to obey velocity limits.
Here, it is non-trivial to predict when such a~switch must happen
when modeling dynamics based on differential equations.

\full{%
We show that it is possible to~solve specific tasks efficiently
even with a~general purpose algorithm.
The only parts that are dedicated to the particular railway scheduling problem
are an~appropriate decision heuristic,
and the way how a~formula is formed.
}%

\full{%
\paragraph{Related work.}
}%
We are not aware of any other approach to railway scheduling
based on SAT modulo theories with realistic modeling of continuous dynamics.
The mentioned approach~\cite{rail-art} solves the problem of design capacity analysis,
a different, but related problem. The main differences are:
\begin{itemize}
\item Instead of an~ad-hoc combination of SAT and a~simulator,
    we model the problem in a~precisely defined \acrl{smt} language\open{~\cite{my-tap-art}}.
    As a~result, numeric (e.g. timing) constraints can appear throughout a~formula.
    \full{%
    They are analyzed in tight integration of the simulator
    and the SAT solver.
    }
\item Our model allows rich timing constraints\short{, including their Boolean combinations}.
    \full{%
    Boolean combinations
    of both relative and absolute timing constraints,
    with upper or lower bounds, are possible.
    }%
    Consequently, trains are allowed to keep waiting
    in stations, or before entering the network,
    even in cases when their routes do not collide with the other trains.
    Hence, our model may exhibit more nondeterminism
    which makes the scheduling problem more difficult.
\item The dynamics of trains is an~integral, but modifiable part of the model,
    instead of being hidden in a~simulator.
\end{itemize}
Both approaches have different strong and weak aspects of the run-time performance.

As in the case of any formal model of real-world problems,
also here, we abstract from certain aspects of the problem domain.
Our model does not take into account railway policies\full[, ]{
or such,
and our approach may be more generic than it is actually necessary in practice.
For example, we do not model objects like
}%
signaling principles\full{, train detectors, switches}, and the like,
as Luteberget et. al.~\cite{rail-art} do.
%% It can be non-trivial to layout the signals and detectors properly
%% In the end, it has to be there (for regular railways), and from our output,
%% it is probably not clear where to place them,
%% or if it is consistent with an existing layout of signals
Especially we do not claim \acrfr{etcs} compatibility of our model,
meaning that it may be less suitable for railway systems
based on signal interlocking.
However, not all railways use such a~mechanism,
for example urban railways may leave the responsibility to the driver\hide{, instead}.
\full{%
More specific comparison of particular differences
follows in related sections.
}

Railway route planning can also be viewed
as a~multi-agent path finding problem~\cite{rail-mapf-art},
where trains are viewed as agents.
However, in this area,
usually much simpler models of continuous behavior are used~\cite{multi-agent-art}.
On the other hand, the resulting plans are often minimized
wrt. a~given parameter, for example, sum of lengths of the agents' paths,
while we do not optimize at all.

Of course, many other approaches dedicated to railway scheduling exist.
Some support only limited precision,
or work only under certain assumptions, for example, fixed routes,
or not taking into account limited track capacity.
Some use networks that were transformed from a~microscopic level\fullfootnote{%
Microscopic level corresponds to railway simulation, like in our case.
}
to an~aggregated, macroscopic level~\cite{rail-networks-art}.
\full{%
Some use an~approximation where both microscopic and macroscopic models
are included~\cite{rail-sched-micro-macro}.
}%
Also, probabilistic methods exist~\cite{rail-sched-prob,rail-sched-prob-macro}.

There are approaches\full{~\cite{rail-sched-art,rail-sched-fixed-timetables-art}}
that are quite accurate,
but still ignore some constraints that we take into account\full[.
For example, not all combinations of possible train paths are considered~\cite{rail-sched-art},
or bi-directional tracks are replaced by pairs of one-directional tracks,
and simpler train dynamics is used~\cite{rail-sched-fixed-timetables-art}.
]{,
for example:
\begin{itemize}
\item Not all combinations of possible train paths are considered~\cite{rail-sched-art}.
\item They are based on an~already existing time table,
    which can even be assumed to be fixed~\cite{rail-sched-fixed-timetables-art}.
    No such prerequisites are necessary in our case.
\item They still over-approximate the available
    capacity~\cite{rail-sched-fixed-timetables-art},
    by replacing bi-directional tracks by pairs of one-directional tracks.
    In our case, the topology of an~infrastructure
    can correspond to the reality accurately.
\item Simpler train dynamics is used~\cite{rail-sched-fixed-timetables-art}.
\end{itemize}
On the other hand, most of the approaches that were mentioned so far
are optimization algorithms,
while we present a~decision procedure.
}

The paper is structured as follows.
We start with an explanation of the problem area in \refSec{s:problem}.
We briefly describe the used theory in \refSec{s:theory},
and present encoding of the problem as a~formula of that theory in \refSec{s:phi}.
The algorithm we use to solve the problem follows in \refSec{s:alg}.
Finally, in \refSec{s:exp}, we analyze the behavior of our approach and of~\cite{rail-art}
on selected case studies.
\short{%
Some parts of the study are omitted due to lack of space,
but are available {\open[in appendix]{%
in an~extended version of the paper~\cite{my-rail-online-art}%
}}.}

%%%%%%%%%%%%%%%%%%%%%%%%%%%%%%%%%%%%%%%%%%%%%%%%%%%%%%%%%

\section{Problem Overview}\label{s:problem}
\full{%
The problem is related to finding a~low-level schedule
within a~railway network.
In other words, it is \emph{searching} for a~simulation of trains
that meets given properties.
The real-world objects, like trains and railroads,
are being abstracted in the form of mathematical models.
}

This section describes the overall problem and introduces related keywords.
We start with an~illustrative example.

%%%%%%%%%%%%%%%%%%%%%%%%%%%%

\subsection{Example}\label{ss:problem:example}
\Img{network}
    {An~example of a~rail network graph with trains}
    {\IncludeImg{0.7}{railway}}

In \refImg{network},
one can see a~model of a~rail network with three trains.
We distinguish the model itself and the required constraints on trains.

\paragraph{The model} consists of a~graph of the network,
and of abstracted trains.
Each train is described by its physical properties,
for example length, velocity limit, etc.
The red train is a~freight train, longer and slower than the other,
passenger trains.
For illustrative reasons,
the boundary nodes of the graph are distinguished from the others.
The thicker an~edge is, the faster railroad it represents.
Nodes that model stations are labeled with a~number.
To support modeling of railway junctions,
nodes of the graph have two sides, illustrated by black and blue colors in the figure.
In order to avoid physically impossible (e.g. too sharp) turns,
a~train has to visit both sides when transferring via such a~double-sided node.

\paragraph{Constraints.}
Examples of constraints, that the trains in the figure might be required to satisfy, are:
\begin{itemize}
\item The blue train must start from the boundary~A,
    and has no further requirements on visiting nodes.
\item The green train must start from~A,
    and is in addition required to visit node~3,
    where it will stop.
    Eventually, the train must continue to node~D afterwards.
\item The red train must start at~D and exit at~A, with no other required visits.
\item Possible orderings of the trains:
    the blue train
    must start before the green train;
    the red train starts before the green train approaches node~1.
\item Possible timings of the trains:
    the red train must arrive at~A within 10~minutes after entering;
    the green train must wait at node~3 for at least 2~minutes.
\end{itemize}

\paragraph{The result} of the search is a~plan
that demonstrates how the trains can move through the network
while satisfying the given constraints and with no collisions of trains.
\full{%
An~example of such a~plan as a~whole follows.

The blue train entered first and is currently approaching the boundary~B,
not interfering the other two trains.
The green train is free to enter next,
but the red train must enter before the green approaches~1.
The red could also enter before the green,
which could enter as late as the red exits,
but we assume the case that the green enters before the red.

So, the green train aims to the node~3,
and to exiting the network at the boundary~D afterwards
(it could also be decided that it would keep waiting at the node~3).
However, as soon as the red train enters,
the collision of the trains must be avoided,
so the red has to bypass the node~1 via the node~2,
and then aim to the boundary~A.

This way, all the mentioned constraints were met,
and the resulting schedule is also efficient
wrt. the railway capacity and elapsed real-time.
(Of course, it could also ended up with a~not so efficient one.)
}

%%%%%%%%%%%%%%%%%%%%%%%%%%%%

\subsection{General Problem Statement}\label{ss:problem:general}
The task is to find a~plan for a~given set of trains
and a~railway network (viewed as a~graph)
such that all specified places are visited,
meeting all timing and ordering constraints, and with no collisions of trains.
\full{%
It is a~\emph{satisfiability} problem---it \emph{decides} whether there exists
an~assignment of variables that satisfies a~formula.
If it exists, the assignment is supplied,
optionally including the trajectories of trains.
}

We assume that each train can only enter the network at a~boundary,
and that at the beginning of the whole search,
there are no trains present in the network.
Also, trains are not allowed to reverse their direction.
\full{%
All these aspects can be included in the model, though.
}
%%+ to natively support terminus stations, add sth. better than 'wait(T, N) > inf)'
%%+ (which would be in conflict with FINISHED, though)

\full{%
The decisions to be made for each train are
branching edges, when to enter the network,
and if it stopped at a~station, when to exit it;
everything else is, ideally, deterministic.
}

%%%%%%%%%%%%%%%%%%%%%%%%%%%%

\subsection{Railway Model}\label{ss:problem:model}
\full{%
A~closer specification of what we support in the model,
for example, which properties of trains,
or which kind of graphs, follows.

The environment of the presented experiments
is a~low-level model of railway transport.
It consists of a~steady infrastructure,
that can actually relate to real world railroads,
and of a~given number of various train models
with realistic dynamic behavior.
}

%%%%%%%%%%%%%%

\subsubsection{Infrastructure.}\label{sss:problem:model:infra}
An~infrastructure (or a~network)
is modeled using a~graph of vertices called \emph{nodes} and edges called \emph{segments}.
Each segment has a~length and a~velocity limit.
\full{%
Properties like segment's slope, angle or cant, or whether it is a~tunnel,
are missing.
}%
Only a~single train is allowed inside a~segment
and the chosen next segment where the train currently aims to.
A~node that is not boundary
either may or may not allow stopping,
where nodes that allow stopping model stations.
The fact that a~train shall stop at a~node
is not modeled explicitly, but by temporarily setting
the velocity limit of the train's chosen next segment
(which the other trains are not allowed to enter) to zero.
After stopping, trains may wait in stations for a~limited time, or may not.

As explained in the example (\refSec{ss:problem:example}),
the graph is a~\emph{double-vertex} graph~\cite{double-vertex-art},
which is commonly used for modeling railways with junctions~\cite{rail-networks-art}.

\hide{%
We actually used an~equivalent model with single nodes
but with two additional special types of edges,
which correspond to which sides of neighboring nodes
in the double-vertex graph are being connected.
The motivation behind this is that in most cases,
the basic edge is sufficient to use,
and the special edges are necessary only rarely.
}

We assume that each segment is at least as long as the longest train
(\refImg{network} violates this property).
\full{%
As a~result, each train is always present within at most two segments on its way.
It should not be difficult to improve the encoding
s.t. this restriction is not necessary, though.
\hide{%
Also, segments should be
as long as the distance required by braking of any train
from the velocity limit of the segment
until the velocity limit of any neighboring segment,
otherwise the plan will be unsatisfiable.
}}%
The model directly supports infrastructures with cycles
and looping of trains,
in contrast with~\cite{rail-art} where this needs an~extra effort.

%%%%%%%%%%%%%%

\subsubsection{Train.}\label{sss:problem:model:train}
A~train~$T$ has an~acceleration and a~deceleration rate,
a~velocity limit, and a~length.
The dynamics of trains is deterministic---%
each train drives at the maximum possible speed,
which, however, depends on discrete decisions---%
the choice of segments on the train's way,
and where to stop.
Such a~model already allows meaningful experiments,
but can be easily extended\full[.]{
by features like weight,
number of wagons, etc.
}

%%%%%%%%%%%%%%%%%%%%%%%%%%%%

\subsection{Constraints}\label{ss:problem:const}

%%%%%%%%%%%%%%

\subsubsection{Connection Constraints.}\label{sss:problem:const:conn}
A~\emph{connection} is a~mapping of a~train to a~non-empty list of nodes
that must be visited in the given order.
For instance, \(T_\var{green} \mapsto (A,3)\)
is the connection
of the green train from the example.
The user must specify exactly one connection for each train.
The list can contain boundary nodes too, but only as the first or the last element.
The first element of the list indeed must be a~boundary node.
Trains always stop at the listed nodes
that model stations,
and never stop at any other stations.
\full{%
Other attitudes can be considered too
(e.g., trains do stop there too, or they \emph{may} stop there).
}%
\full{%
For example, regarding the connection \(\var{green} \mapsto (A,3)\),
the green train will stop at node~3, but will not stop at node~1, because it is not listed,
and even if the list contained the node between nodes~A and~1,
the train will not stop there neither, because it does not model a~station.
}

The \emph{starting} node is the first node in the list.
A~connection may have several \emph{ending} nodes---%
any boundary node
terminating a~path following the given connection.
For example, in \refImg{network}, given a~connection list \((A,2)\),
$A$ is the starting node and $C,D$ are two possible ending nodes,
but for connection list \((A,2,D)\), $D$ is the only ending node.
We call segments incident with the starting node \emph{starting segments},
and segments incident with an~ending node \emph{ending segments}.

%%%%%%%%%%%%%%

\subsubsection{Schedule Constraints.}\label{sss:problem:const:sched}
Schedule constraints are optional constraints
that compare the time when a~train either arrives at or departs
from a~node\fullfootnote{%
For a~required visit, it can be useful to also
support specifying sets of nodes,
instead of single nodes (meaning ``any of the nodes''),
which is possible to encode into the formula,
but we do not support such a~rule at the moment.
}.
\full{%
A~departure is when a~train starts accelerating to~leave a~node
after it stopped there earlier, or when entering the network.
\par
}%
In the following, we will denote by $\var{arrival}(T,N)$ (or~$\var{departure}(T,N)$)
the time when train~$T$ arrives at (or departs from) node~$N$.
To~allow both variants in a~formula,
we will write $\var{visit}(T,N)$,
possibly distinguishing several occurrences by indices
($\var{visit}_1(T_1,N_1)$, $\var{visit}_2(T_2,N_2)$, etc.).
Schedule constraints assume that all mentioned visits are the consequence
of some connection constraint.

We allow two types of schedule constraints, ordering, and timing constraints.
An~\emph{ordering} enforces two visits to happen in a~given order.
It has the form
\begin{equation}\label{\eqL{problem:const:sched:order}}
\var{visit}_1(T_1, N_1) \circ \var{visit}_2(T_2, N_2)
,
\end{equation}
where $\circ$ is one
of~\full[$\{ <, \leq \}$]{%
$\{ <, \leq, >, \geq, = \}$.
In the case of \cite{rail-art},
the only supported orderings are with $\circ$ being~$<$ or~$>$%
}.
A~\emph{\short{relative }timing}
\full[enforces ]{%
is either absolute, concerning one visit:
\begin{equation}\label{\eqL{problem:const:sched:time:abs}}
\var{visit}(T, N) \circ \xi
,
\end{equation}
or relative, requiring
}%
a time constraint on a transfer, that is, on the time from one visit to another.
It has the form
\begin{equation}\label{\eqL{problem:const:sched:time:rel}}
\var{transfer}(\var{visit}_1(T_1, N_1), \var{visit}_2(T_2, N_2)) \circ \xi
,
\end{equation}
where
\(\var{transfer}(v_1, v_2) \coloneqq v_2-v_1\),
\(\circ \in \{ <, \leq, >, \geq \}\),
and \(\xi \in \numset{Q}_{\geq 0}\).
\full{%
Also,
\(\var{wait}(T, N) \coloneqq \var{transfer}(\var{arrival}(T, N), \var{departure}(T, N))\).
}%
\short{%
We support absolute timings, as well, but they are omitted here.
}%
In the case of~\cite{rail-art},
the only supported timing constraints are
\(\var{transfer}(\var{arrival}(T_1, N_1), \var{arrival}(T_2, N_2)) < \xi\)\full[.]{,
which is quite sufficient
for the purposes of railway capacity verification, though.
}

%%%%%%%%%%%%%%%%%%%%%%%%%%%%%%%%%%%%%%%%%%%%%%%%%%%%%%%%%

\section{Theory Description}\label{s:theory}
For encoding our problem, we use a~\acrf{smt}~\cite{smt-dpll_t} language.
For this, we use a~theory for reasoning about \acrl{ode}s (\acr{ode}s)
that \open[was]{we} introduced earlier~\cite{my-tap-art}.
In this section, we provide an~informal summary of this theory.

In addition to variables ranging over the real numbers,
together with the usual operations on them,
the theory allows variables ranging over real functions
\(\interval{0}{\tau} \rightarrow \numset{R}\),
that \open[are called]{we call} \emph{functional variables}.
Here \(\tau \in \numset{R}^{\geq 0}\).
Functional variables can be constrained
by \emph{differential constraints} of the form \(\der{x}=\eta\)
with \fun{x} being a~functional variable
and $\eta$ a~term containing functional and real variables;
by \emph{invariants} that have to hold over the whole interval~\interval{0}{\tau},
and by real-valued constraints that restrict the initial value \var{init}
or final value \var{final} of a~functional variable.
For example, the formula
\(\var{init}(\fun{x}) = 0 \land \der{x} = \fun{x} \land \fun{x} \leq 10\)
restricts the initial value of the functional variable \fun{x} to~zero,
restricts its evolution over time by the differential equation \(\der{x} = \fun{x}\),
and restricts \fun{x} to functions for which the upper bound~$\tau$
of the time interval is such that
the invariant \(\fun{x} \leq 10\) holds over the whole interval.

Since this theory is undecidable~\cite{continuous-survey},
\open{we also introduced }an~alternative semantics
that approximates the mathematical semantics
using floating-point numbers~\cite{float-art} and simulations of ODEs
(i.e. numeric integrations)\blind{ was also introduced in~\cite{my-tap-art}}.
Using this semantics, the theory is not only decidable in the theoretical sense,
but also efficiently decidable for formulas
of the type occurring in this paper.
We will use this semantics in the algorithm in \refSec{s:alg}
and in all our experiments\full{, as described} in \refSec{s:exp}.

In contrast to the original semantics~\cite{my-tap-art},
where simulations could be terminated before violation of an invariant,
here we always continue simulations until an~invariant is violated\full{,
without expressing this explicitly with the \var{final} constraints%
}.
As a~consequence the length of each interval~\interval{0}{\tau} is completely determined
and all nondeterminism in the model described below
will stem from discrete decisions.
\full{%
For example, although one requires interval timing constraints,
this does not mean that all times from the interval will be tried,
only all discrete consequences that lead inside this interval.
}

%%%%%%%%%%%%%%%%%%%%%%%%%%%%%%%%%%%%%%%%%%%%%%%%%%%%%%%%%

\section{Encoding and Formalization}\label{s:phi}
In this section, we present an~encoding
of the planning problem from \refSec{s:problem} as a~formula
in the theory described in \refSec{s:theory}\full{
(but the floating-point semantics are not necessary here)%
}.
All of the presented formulas are generated automatically,
from user input in the form of a~preprocessing
language~\opencite{unsot-preprocess_lang}\full{%
to a~format~\opencite{unsot-core_lang}
that roughly follows the SMT-LIB format~\cite{smtlib-reference-art}%
}.
The user input consists of specification of an~infrastructure and of trains,
and of connections and schedule constraints.

We unroll the planning problem in a~similar way as in \acrf{bmc}~\cite{bmc-art}.
Unrolling ranges over discrete steps $0, 1, \dots, J$.
A~variable~$x$ specific to a~discrete step~$j$
has the form~\atj{x}.
All trains are modeled \emph{synchronously},
meaning that every discrete step~$j$
corresponds to the same global moment in time.
Functional variables specific to one and the same discrete step
will have the same length~\atj{\tau} of integration,
from which we get global time by defining
real variables~\atj{t} s.t. \(\at{t}{0} = 0\)
and for all \(j > 0\), \(\atj{t} = \at{t}{j-1} + \at{\tau}{j-1}\).
\full{%
This simplifies the encoding and analysis,
because it is simple to~compare current values of trains' variables.
This implies that all switches from a~discrete step $j$ to $j+1$ are globally shared.
However, synchronicity also implies
that to~cover overall dynamics of all trains,
the total number of discrete steps must be quite high,
which has a~bad influence on performance.
}%
In the case of~\cite{rail-art},
the planner considers longer units for unrolling
where a~step may consist of movements over several segments,
and within such a~step all deterministic discrete constraints are handled by the simulator.

%% Many elementary routes may be alloc. to a train in a single planning step (!)
%% we need e.g. discrete step for every velocity limit change, not their case ...

We use one-hot encoding for some Boolean variables
for increased readability.
Moreover, we improve readability by using an~abbreviation
\ite{\var{cond}}{a}{b}
for~\(\bigl((\var{cond} \Rightarrow a) \land (\neg \var{cond} \Rightarrow b)\bigr)\).

\blind{%
Some parts omitted here due to lack of space are available in \refAppend{a:phi}.
}

%%%%%%%%%%%%%%%%%%%%%%%%%%%%

\subsection{Railway Model}\label{ss:phi:model}
\full{%
The model itself is represented by a~formula entirely.
Almost all variables are related to a~particular train~$T$,
including the graph representation.
The only variables that are global
are those related to the shared time.
}

\short{%
We only present the most significant constraints
that are necessary for understanding the principles of the model.
See {\open[\refAppend{aa:phi:model}]{%
the extended version of the paper~\cite{my-rail-online-art}
}}%
for more details.
}

%%%%%%%%%%%%%%

\full{%
\subsubsection{Infrastructure.}\label{sss:phi:model:infra}
The graph that models the network is not itself represented
by dedicated variables, but instead,
by Boolean variables and constraints
related to each particular train~$T$,
which define possible transitions of the train.
All segments \(S \in \set{S}\) and nodes \(N \in \set{N}\)
of the graph
are just identifiers that serve as parts of the variable names
related to the train.
Incidency of all the nodes and segments
is covered within the preprocessing stage,
which is when the corresponding Boolean variables and constraints
are generated into the formula---see below.
}

%%%%%%%%%%%%%%

\subsubsection{Train.}\label{sss:phi:model:train}
A~train \(T \in \set{T}\) is defined by fixed constants
\xth{A}{T},~\xth{B}{T},~\xth{\var{V_{max}}}{T}, and~\xth{L}{T}
that represent the properties of the train
(acceleration\full[,]{ and} deceleration\full{ rate}, velocity limit and length)%
\full[;]{, where}~$T$ is just a~prefix of the constant names,
representing an~identifier of the train.
In a~similar way,
the state of each train is described by a~set of variables,
distinguished by a~discrete step~$j$.
The most important variables are:
\begin{itemize}
\item \emph{Booleans}:
    \atj{\xth{\var{mode}}{T}}, \(\var{mode} \!\in\! \set{M} =
         \{\var{idle}, \var{steady}, \var{acc}, \var{brake}\}\)
        (\var{steady} means the train does not accelerate,
        but in mode \var{idle}, in~addition, it has zero velocity);
    \atj{\xth{\var{away}}{T}},~\atj{\xth{\var{enter}}{T}} and~\atj{\xth{\var{finished}}{T}}
        (whether the train is currently outside the graph,
        whether it is entering, and whether it already finished);
    and~\atj{\xth{\var{pos\_S}}{T}},
    \(\var{pos} \!\in\! \set{P} =
        \{\var{back}, \var{front}, \var{next}\}\),
        for a~segment \(S \!\in\! \set{S}\)
        (the train's back and front being in~$S$;
        whether $S$ is selected as the next segment).
\item \emph{Reals}:
    \atj{\xth{a}{T}} (acceleration/deceleration rate);
    \atj{\xth{\var{d_{max}}}{T}}
        (remaining distance to the end of the current segments,
        either with the back or the front of the train,
        i.e., for segments~$S$ where
        \atj{\xth{\var{back\_S}}{T}} or \atj{\xth{\var{front\_S}}{T}} holds);
    \atj{\xth{\var{v_{max}}}{T}}
        (velocity limit of the current segments and the train itself);
    and~\atj{\xth{\var{next\_v_{max}}}{T}}
        (velocity limit of the selected next segment~$S$,
        for which \atj{\xth{\var{next\_S}}{T}} holds).
\item \emph{Functional variables}:
    \atj{\xth{\fun{d}}{T}}, \(\var{init}(\atj{\xth{\fun{d}}{T}}) = 0\)
        (relative distance traveled from the start of unrolling step~$j$),
    and~\atj{\xth{\fun{v}}{T}} (current velocity).
    The functional variables
    range over~\interval{0}{\atj{\tau}},
    where a~timeout \(\atj{\tau} < \rho\),
    with the constant $\rho$ user-defined,
    must hold.
    This allows decisions on when to enter the network
    or when to leave the current station
    to happen in certain intervals---%
    if the timeout is too short,
    the number of necessary discrete steps may be too high;
    if it is too long, a~plan where trains stay idle for too long
    may be returned.
\end{itemize}

%%%%%%%%%%%%%%

\subsubsection{Result.}\label{sss:phi:model:res}
The resulting plan is represented by the global variables \atj{t}
and the variables \atj{\xth{\var{idle}}{T}}, \atj{\xth{\var{front\_S}}{T}}
and~\atj{\xth{\var{finished}}{T}},
for all trains $T$, segments $S$ and discrete steps $j$.
All other variables are either auxiliary or are completely determined
by the plan and the model described in this subsection.

%%%%%%%%%%%%%%

\subsubsection{Dynamic Phenomena.}\label{sss:phi:model:dyn}

\paragraph{Mode conditions.}
Unlike in capacity analysis~\cite{rail-art},
where behavior is deterministic,
as soon as routes have been chosen,
here continuous dynamics depends on each train's mode,
where a~train can choose to stay idle in stations, or before entering the network.
Each train~$T$ is always in exactly one dynamic mode\full[, ]{:
\begin{equation}\label{\eqL{phi:model:mode}}
\bigvee_{\var{mode} \in \set{M}} \atj{\xth{\var{mode}}{T}}
\ \land\
\bigwedge_{\var{mode}_1, \var{mode}_2 \in \set{M}, \var{mode}_1 \neq \var{mode}_2} \neg\bigl(
    \atj{\xth{\var{mode}_1}{T}} \land \atj{\xth{\var{mode}_2}{T}} \bigr)
\end{equation}
}%
and according to this mode, an~appropriate (constant) acceleration rate is set:
\begin{equation}\label{\eqL{phi:model:mode:a}}
\begin{split}
\bigl( (\atj{\xth{\var{idle}}{T}} \lor \atj{\xth{\var{steady}}{T}})
    \Leftrightarrow \atj{\xth{a}{T}} = 0 \bigr)&
\\\land\ \bigl(
\atj{\xth{\var{acc}}{T}}
    \Leftrightarrow \atj{\xth{a}{T}} = \xth{A}{T} \bigr)&
\land \bigl(
\atj{\xth{\var{brake}}{T}}
    \Leftrightarrow \atj{\xth{a}{T}} = -\xth{B}{T} \bigr)
.
\end{split}
\end{equation}

\newcommand{\TextDynamicsSwitchMode}{%
To~reduce nondeterminism of switching of the modes, we add:
\begin{equation}\label{\eqL{phi:model:mode:jump}}
\begin{split}
\Bigl( \atj{\xth{\var{idle}}{T}} &\Rightarrow
    \bigl(\at{\xth{\var{idle}}{T}}{j+1} \lor \at{\xth{\var{acc}}{T}}{j+1}\bigr) \Bigr)
\\\land\ \Bigl(
    \bigl(\atj{\xth{\var{steady}}{T}} \lor \atj{\xth{\var{acc}}{T}}\bigr)
        &\Rightarrow \bigl(\neg\at{\xth{\var{idle}}{T}}{j+1} \lor
            \at{\xth{\var{away}}{T}}{j+1}\bigr) \Bigr)
.
\end{split}
\end{equation}
Furthermore, sometimes it is clear that braking must follow\full{
(i.e. constraints on~\at{\xth{\var{brake}}{T}}{j+1}),
but it is discussed later, within the paragraph with braking prediction%
}.
}

\full{\TextDynamicsSwitchMode}

There are also other restrictions, like
that braking is not possible if the velocity is already zero,
or that \var{steady} mode is not allowed if acceleration is possible.

\Img[tb]{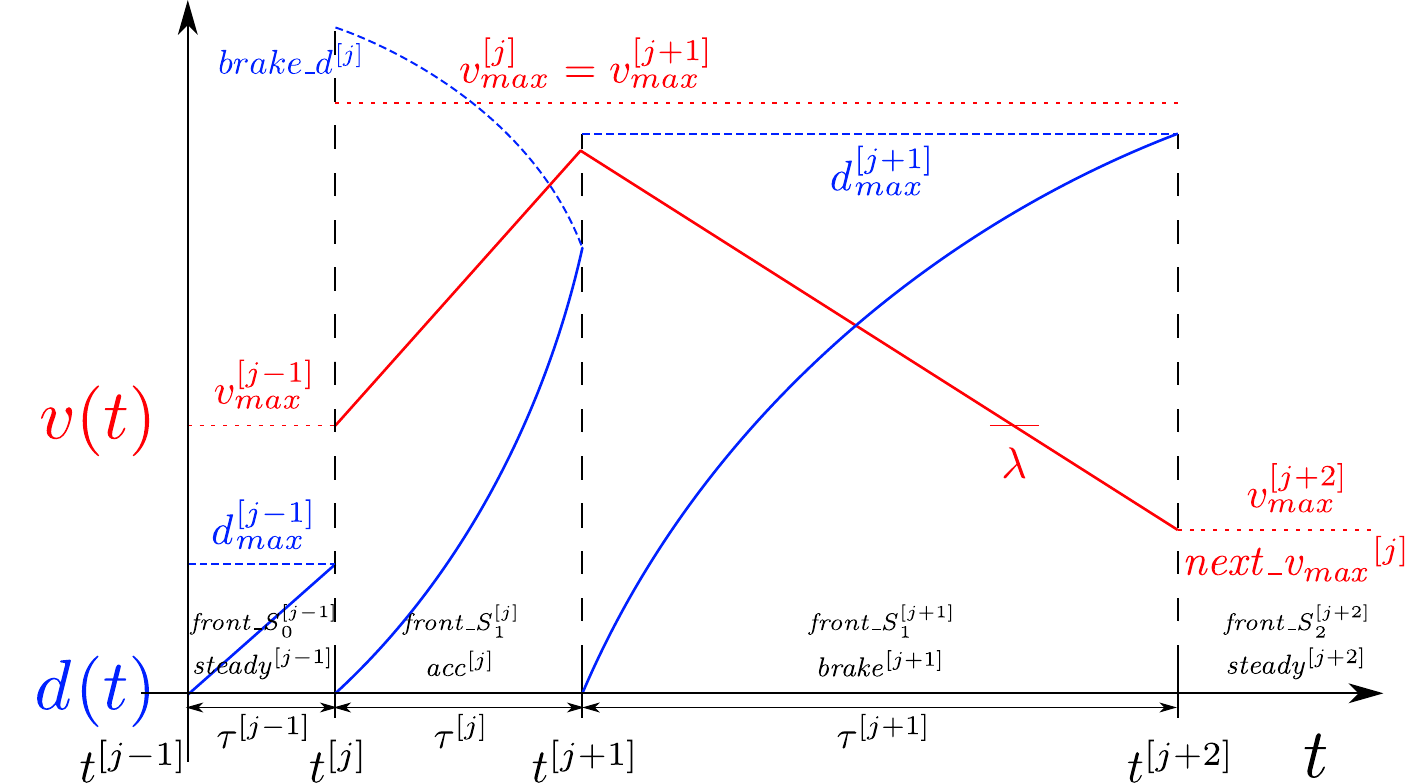}
{Possible train trajectories and their limits}
{\IncludeImg{0.8}{train_brake}}

\paragraph{Dynamics.}
We model the dynamics of trains using the basic laws of motion,
but it is possible to extend the model
such that it exhibits more complex phenomena\full[. ]{,
like engine power curves, tunnel air resistance, curve rolling resistance,
train weight distribution, etc.
}%
\refImg{train_brake} illustrates
how the resulting trajectories of functional variables can look like
($T$~is omitted from the variable names).
Both functions~\fun{v} and~\fun{d} are limited
by a~corresponding dashed line,
a~constant \at{\var{v_{max}}}{\cdot} in the case of the function~\fun{v},
and a~distance limit in the case of~\fun{d},
either in the form of a~straight line, representing \at{\var{d_{max}}}{\cdot},
or a~curve, that stands for the function \at{\fun{brake\_d}}{\cdot}
that is about to be discussed further.
The limit \at{\var{v_{max}}}{j+2}
is equivalent to~\atj{\var{next\_v_{max}}}\full{
(and to~\at{\var{next\_v_{max}}}{j+1} as well)%
}.
\full{%
In the figure, it is always a~distance limit that ends each stage,
because no velocity limit is exceeded there.
}

Since trains are modeled synchronously,
the dynamics of the trains
is represented mainly by one system of \acr{ode}s---%
for each train~$T$, and discrete step~$j$:
\begin{equation}\label{\eqL{phi:model:dyn}}
\begin{split}
\atj{\xth{\der{d}}{T}} = \atj{\xth{\fun{v}}{T}}
    \quad&\land\quad
        \atj{\xth{\der{v}}{T}}  = \atj{\xth{a}{T}}
\\\land\quad
\atj{\xth{\fun{d}}{T}} \leq \atj{\xth{\var{d_{max}}}{T}}
    \quad&\land\quad
        \atj{\xth{\fun{v}}{T}} \in \interval[{\bigl[}][{\bigr]}]{0}{\atj{\xth{\var{v_{max}}}{T}}}
.
\end{split}
\end{equation}
The first row of the formula shows particular ODEs,
and the second the invariants.
Thus, each integration ends when
a~distance limit or a~velocity limit is exceeded,
or when the timeout is reached,
which was explained in the description of functional variables.

For the definition of the variables
\atj{\xth{\var{\alpha_{max}}}{T}}, \(\alpha \in \{\fun{d}, \fun{v}\}\),
we use auxiliary variables
\atj{\xth{\var{pos\_\alpha_{max}}}{T}},
\(\var{pos} \in \set{P}\)
which correspond to the limits of the current and the next segments,
as mentioned in the description of the real variables.
Moreover, \(\atj{\xth{\var{min\_\alpha_{max}}}{T}} \coloneqq
\min\{\atj{\xth{\var{back\_\alpha_{max}}}{T}}, \atj{\xth{\var{front\_\alpha_{max}}}{T}}\}\).
Then, the distance limit is defined by
\(\atj{\xth{\var{d_{max}}}{T}} = \atj{\xth{\var{min\_d_{max}}}{T}}\)
and the velocity limit as
\begin{equation}\label{\eqL{phi:model:dyn:v_max}}
\begin{split}
\ite[\bigl(][\bigr)]
    {&\var{init}(\atj{\xth{\fun{v}}{T}}) \geq \atj{\xth{\var{next\_v_{max}}}{T}}}
    {\\&\atj{\xth{\var{v_{max}}}{T}} = \min\{ \xth{\var{V_{max}}}{T}, \atj{\xth{\var{min\_v_{max}}}{T}} \}}
    {\\&\atj{\xth{\var{v_{max}}}{T}} = \min\{ \xth{\var{V_{max}}}{T}, \atj{\xth{\var{min\_v_{max}}}{T}},
        \atj{\xth{\var{next\_v_{max}}}{T}} \}}
,
\end{split}
\end{equation}
where \atj{\xth{\var{next\_v_{max}}}{T}}
is used to ensure correctness of braking prediction.

\paragraph{Braking prediction.}
In \refImg{train_brake}, within stage~$j$,
one can see that the function \fun{d} is limited
by a~yet unexplained function \atj{\fun{brake\_d}}.
Such a~function is necessary for prediction of the moment
when a~train has to start braking
to obey the velocity limit of the next segment---%
in cases when \(\atj{\xth{\fun{v}}{T}} > \atj{\xth{\var{next\_v_{max}}}{T}}\)
(if the train is not already braking).
The main idea is to compute the braking trajectory backward
from the point where the train enters the next segment,
synchronously with the actual forward dynamics.
Details follow.

The prediction depends on the relation
\(\var{init}(\atj{\xth{\fun{v}}{T}}) \circ \atj{\xth{\var{next\_v_{max}}}{T}}\),
where \(\circ \in \{=, >\}\).
First, let us assume that
\(\var{init}(\atj{\xth{\fun{v}}{T}}) = \atj{\xth{\var{next\_v_{max}}}{T}}\).
To~make \(\atj{\xth{\fun{v}}{T}} > \atj{\xth{\var{next\_v_{max}}}{T}}\) happen eventually,
\atj{\xth{\var{acc}}{T}} must hold.
Such a~case would correspond to \refImg{train_brake},
if \atj{\var{next\_v_{max}}} was in the place of the separator~$\lambda$.
Since \atj{\xth{a}{T}} from \refEq{phi:model:dyn} is a~constant
(due to \refEq{phi:model:mode:a}),
the ratio between the length (in time) of the acceleration phase and the braking
phase is fixed.
Since the temporal relationship between the two phases is not yet clear,
we use independent time axes, writing
\begin{equation}\label{\eqL{phi:model:dyn:time_axes}}
\frac{dv_A}{dt_A} = \xth{A}{T} , \frac{dv_B}{dt_B} = -\xth{B}{T}
,
\end{equation}
where $v_A$ and~$v_B$ corresponds to~\atj{\xth{\fun{v}}{T}} and~\at{\xth{\fun{v}}{T}}{j+1}, resp.,
and~$t_A$ and~$t_B$ corresponds to~\atj{\tau} and~\at{\tau}{j+1}, resp.
To~determine the time to switch from acceleration to braking,
it would be possible to compute the braking trajectory backward in time
starting at the position corresponding to~\atj{\xth{\var{front\_d_{max}}}{T}},
and with the velocity corresponding to~\atj{\xth{\var{next\_v_{max}}}{T}}.
However, it is not clear how far backward such a~backward braking
trajectory has to be computed, and moreover, even after its computation,
it is non-trivial to ensure that at the switching time,
\emph{both} position and velocity of the train are identical
to a~corresponding point on the backward braking trajectory.
To get around these complications, we not only reverse,
but also scale the time axis of the braking process using the relationship
\begin{equation}\label{\eqL{phi:model:dyn:scale}}
t_B = -\frac{\xth{A}{T}}{\xth{B}{T}} \cdot t_A
.
\end{equation}
As a~result, we have a~common time axis~$t_A$,
along which the derivative of the velocity of the braking train
is identical to the derivative of the velocity of the accelerating train:
\begin{equation}\label{\eqL{phi:model:dyn:time_axis:common}}
\frac{dv_B}{dt_A} = \frac{dv_B}{dt_B} \frac{dt_B}{dt_A} =
-\frac{dv_B}{dt_B} \frac{\xth{A}{T}}{\xth{B}{T}} =
\xth{B}{T} \cdot \frac{\xth{A}{T}}{\xth{B}{T}} =
\frac{dv_A}{dt_A}
.
\end{equation}
As a~consequence, both velocities will be identical at all time
if starting from the same initial value.
Under this assumption, we can compute both the acceleration phase
and the backward braking trajectory synchronously along the same time axis,
ensuring identical speed at all times.
Such an~approach can be generalized for more complicated systems of ODEs
(e.g. with \atj{\xth{\fun{v}}{T}} other than a~linear function),
if such a~relationship between the time axes is available.

Based on \refEq{phi:model:dyn:time_axis:common}, it suffices to~switch from acceleration
to braking at the point when the corresponding positions are identical.
This results in a~synchronous braking prediction
with ODEs and an~invariant
of the form
\begin{equation}\label{\eqL{phi:model:dyn:brake}}
\begin{split}
\ite[\bigl(][\bigr)]{\atj{\xth{\var{acc}}{T}}}
    {\atj{\xth{\der{brake\_d}}{T}} =
    - \frac{\xth{A}{T}}{\xth{B}{T}} \cdot \atj{\xth{\fun{v}}{T}} }
    {\atj{\xth{\der{brake\_d}}{T}} = 0}
\\\land\quad \bigl(
\neg\atj{\xth{\var{brake}}{T}} \Rightarrow
    \atj{\xth{\fun{d}}{T}} \leq \atj{\xth{\fun{brake\_d}}{T}}
\bigr)
\end{split}
\end{equation}
where the coefficient $-\frac{\xth{A}{T}}{\xth{B}{T}}$ implements
the mentioned scaling also for the prediction of the position of the train.
\full{%
Note that the same \atj{\xth{\fun{v}}{T}} as in \refEq{phi:model:dyn} is used.
For example, in the figure,
deceleration must proceed more quickly within the braking prediction.
}

If \(\var{init}(\atj{\xth{\fun{v}}{T}}) > \atj{\xth{\var{next\_v_{max}}}{T}}\),
the part of the braking phase
with \(\at{\xth{\fun{v}}{T}}{j+1} \in
\interval{\atj{\xth{\var{next\_v_{max}}}{T}}}{\var{init}(\atj{\xth{\fun{v}}{T}})}\)
must be precomputed asynchronously.
In the figure, this corresponds to the part from the end of stage~$j+1$
to the separator~$\lambda$ (backwards).
Such an~asynchronous prediction uses the functional variables \fun{back\_d} and \fun{back\_v},
starting from
\begin{equation}\label{\eqL{phi:model:dyn:brake:async:init}}
\var{init}(\atj{\xth{\fun{back\_d}}{T}}) = \atj{\xth{\var{front\_d_{max}}}{T}}
\land \var{init}(\atj{\xth{\fun{back\_v}}{T}}) = \atj{\xth{\var{next\_v_{max}}}{T}}
,
\end{equation}
with a~flow defined by the following ODEs and invariants:
\begin{equation}\label{\eqL{phi:model:dyn:brake:async}}
\begin{split}
\atj{\xth{\der{back\_d}}{T}} = -\atj{\xth{\fun{back\_v}}{T}}
    \quad&\land\quad
        \atj{\xth{\der{back\_v}}{T}} = \xth{B}{T}
\\\land\quad
\atj{\xth{\fun{back\_d}}{T}} \geq 0
    \quad&\land\quad
        \atj{\xth{\fun{back\_v}}{T}} \leq \var{init}(\atj{\xth{\fun{v}}{T}})
.
\end{split}
\end{equation}
These functional variables are the only ones that may have a~different
length~$\tau$ of integration than the other variables (which are synchronous).
The reached position serves for the consecutive synchronous part:
\begin{equation}\label{\eqL{phi:model:dyn:brake:async:final}}
\var{init}(\atj{\xth{\fun{brake\_d}}{T}}) = \var{final}(\atj{\xth{\fun{back\_d}}{T}})
,
\end{equation}
and \refEq{phi:model:dyn:brake} becomes computable then.
This works even in cases when \atj{\xth{\var{steady}}{T}} holds,
where \atj{\xth{\fun{brake\_d}}{T}} just serves as a~constant upper bound
on \atj{\xth{\fun{d}}{T}},
based on the value from \refEq{phi:model:dyn:brake:async:final}.

\newcommand{\TextDynamicsSwitchBrake}{%
Consequently to \short{the braking prediction (}\refEq{phi:model:dyn:brake}\short{)},
switching to the braking mode is defined by
\begin{equation}\label{\eqL{phi:model:dyn:brake_mode}}
\begin{split}
\bigl( \atj{\xth{\var{acc}}{T}} &\Rightarrow
    (\at{\xth{\var{brake}}{T}}{j+1} \Leftrightarrow D_j) \bigr)
\\\land\ \bigl( \atj{\xth{\var{steady}}{T}} &\Rightarrow
    (\at{\xth{\var{brake}}{T}}{j+1} \Leftrightarrow ( D_j
        \land \var{init}(\atj{\xth{\fun{v}}{T}}) > \atj{\xth{\var{next\_v_{max}}}{T}} )
    ) \bigr)
\end{split}
\end{equation}
with \(D_j \Leftrightarrow
\var{final}(\atj{\xth{\fun{d}}{T}}) \geq \var{final}(\atj{\xth{\fun{brake\_d}}{T}})\).
If already braking, staying in the mode is defined
depending on not reaching the end of the current segment yet
(see \refEq{phi:model:dyn}),
and possibly also based on the eventual necessity of a~further consecutive braking
(due to \refEq{phi:model:dyn:brake:async}).
That is
\begin{equation}\label{\eqL{phi:model:dyn:keep_brake_mode}}
\begin{split}
\atj{\xth{\var{brake}}{T}} \Rightarrow
    \ite[\bigl(][\bigr)]{&\var{final}(\atj{\xth{\fun{d}}{T}}) < \atj{\xth{\var{d_{max}}}{T}}}
    {\at{\xth{\var{brake}}{T}}{j+1}}
    {\\&\at{\xth{\var{brake}}{T}}{j+1} \Leftrightarrow
        \var{final}(\at{\xth{\fun{back\_d}}{T}}{j+1}) \leq 0}
.
\end{split}
\end{equation}
}

\full{\TextDynamicsSwitchBrake}

If \atj{\xth{\var{v_{max}}}{T}} had been reached before the start of the braking phase\full{
(i.e. the invariant in \refEq{phi:model:dyn} is violated before
the invariant in \refEq{phi:model:dyn:brake})%
},
there just would be an~additional phase in the steady mode
between the phases~$j$ and~$j+1$ in the figure.

\full{%
As a~result, \atj{\xth{\fun{d}}{T}} is limited by both
\atj{\xth{\var{d_{max}}}{T}} (\refEq{phi:model:dyn})
and the braking prediction, which consists of two parts:
the asynchronous part (\refEq{phi:model:dyn:brake:async}),
and the synchronous---\atj{\xth{\fun{brake\_d}}{T}} (\refEq{phi:model:dyn:brake}).
}

%%%%%%%%%%%%%%

\subsubsection{Positional Constraints.}\label{sss:phi:model:pos}
In the following, we use train \(T \in \set{T}\)
and the relation \(S_1 \rightarrow_T S_2\)
for segments \(S_1,S_2 \in \set{S}\)
to~denote that segment~$S_2$ is adjacent to segment~$S_1$
on a~path that obeys the connection constraints of train $T$.
In fact, this relation enforces the connection constraints
completely if \at{\xth{\var{finished}}{T}}{J} (at the final step $J$) holds.
The relation is used only within the preprocessing stage
when generating the formula.

For each segment~$S_1$, the possible next segments are defined s.t.
\begin{equation}\label{\eqL{phi:model:pos:next}}
\neg\atj{\xth{\var{idle}}{T}} \Rightarrow \bigl(
    \atj{\xth{\var{front\_S}_1}{T}} \Rightarrow
        \bigvee_{S_2 \in \set{S}, S_1 \rightarrow_T S_2}
            \atj{\xth{\var{next\_S}_2}{T}}
\bigr)
.
\end{equation}
\newcommand{\TextPosConstraints}{%
\full[Consecutively to \refEq{phi:model:pos:next}, in]{In}
cases when the front of the train is idle,
we do not want to~choose any next segment:
\begin{equation}\label{\eqL{phi:model:pos:not_next}}
\Bigl( \atj{\xth{\var{idle}}{T}}
    \lor \neg\bigvee_{S \in \set{S}} \atj{\xth{\var{front\_S}}{T}} \Bigr)
    \Rightarrow \neg \bigvee_{S \in \set{S}} \atj{\xth{\var{next\_S}}{T}}
.
\end{equation}

Next, a~train cannot be at more segments at once
with any of its part.
That is, for all segments~$S_1$,
\begin{equation}\label{\eqL{phi:model:pos:mutual}}
\bigwedge_{\var{pos} \in \set{P}}\
\bigwedge_{S_2 \in \set{S}, S_2 \neq S_1}
    \neg\Bigl( \atj{\xth{\var{pos\_S}_1}{T}}
        \land \atj{\xth{\var{pos\_S}_2}{T}} \Bigr)
.
\end{equation}
The chosen next segment must be different than the current one:
\begin{equation}\label{\eqL{phi:model:pos:mutual_next}}
\bigwedge_{\var{pos} \in \{\var{back}, \var{front}\}}
    \neg\Bigl(
        \atj{\xth{\var{pos\_S}_1}{T}} \land \atj{\xth{\var{next\_S}_1}{T}} \Bigr)
.
\end{equation}
And, situations like where the train's back is farther than its front are forbidden:
\begin{equation}\label{\eqL{phi:model:pos:order}}
\bigwedge_{\substack{\var{pos}_1 \in \{\var{back}, \var{front}\}
                   \\\var{pos}_2 \in \{\var{front}, \var{next}\}}}
\bigwedge_{S_2 \in \set{S}, S_1 \rightarrow_T S_2}
    \neg\Bigl( \atj{\xth{\var{pos}_2\var{\_S}_1}{T}}
        \land \atj{\xth{\var{pos}_1\var{\_S}_2}{T}} \Bigr)
.
\end{equation}

Finally,
we control the progress of trains in a~way that a~train
must either stay the same, or its back or front has moved forwards
(from a~previous segment),
applied both to the past and to the future.
That is, for all segments~$S$:
\begin{equation}\label{\eqL{phi:model:pos:progress}}
\begin{split}
\neg\atj{\xth{\var{idle}}{T}} \Rightarrow
\bigwedge_{\substack{\var{pos}_1 \in \{\var{back}, \var{front}\}
                   \\\var{pos}_2 = \Delta(\var{pos}_1)}} \Bigl(
    \bigl( \at{\xth{\var{pos}_1\var{\_S}}{T}}{j+1} \Rightarrow (
        \atj{\xth{\var{pos}_1\var{\_S}}{T}} \lor \atj{\xth{\var{pos}_2\var{\_S}}{T}} ) \bigr)
\\\land\
    \bigl( \atj{\xth{\var{pos}_2\var{\_S}}{T}} \Rightarrow (
        \at{\xth{\var{pos}_2\var{\_S}}{T}}{j+1}
            \lor \at{\xth{\var{pos}_1\var{\_S}}{T}}{j+1} ) \bigr)
\Bigr)
,
\end{split}
\end{equation}
with \(\Delta = \{\var{back} \mapsto \var{front}, \var{front} \mapsto \var{next}\}\).
}%
\full{\TextPosConstraints}

\paragraph{Away conditions} distinguish
the cases when a~train already entered the network,
or is outside of it.
The decision variable \var{enter} triggers a~starting segment:
\begin{equation}\label{\eqL{phi:model:enter}}
\atj{\xth{\var{enter}}{T}} \Rightarrow
    \bigvee_{S \in \xth{\var{Start}}{T}}
        \bigl( \neg\atj{\xth{\var{back\_S}}{T}} \land \atj{\xth{\var{front\_S}}{T}} \bigr)
,
\end{equation}
where \xth{\var{Start}}{T} is the set of starting segments of the train~$T$.
\full{%
A~next segment is already constrained by \refEq{phi:model:pos:next}.
After entering, the front of the train is at the beginning
of the chosen segment, while the back is still outside the network,
with the whole train's length.
}%
To denote that a~train is entirely outside the network, we use
\begin{equation}\label{\eqL{phi:model:away}}
\atj{\xth{\var{away}}{T}} \Leftrightarrow
    \neg \Bigl( \bigvee_{S \in \set{S}} \atj{\xth{\var{back\_S}}{T}}
    \lor \bigvee_{S \in \set{S}} \atj{\xth{\var{front\_S}}{T}}
    \Bigr)
.
\end{equation}
The variable \var{finished} is triggered within the transfer constraints
when reaching a~boundary in \refEq{phi:model:transfer:finish} below.
Once the variable is activated, it implies that at least the front of the train is already outside of the network:
\begin{equation}\label{\eqL{phi:model:finished}}
\atj{\xth{\var{finished}}{T}} \Rightarrow
    \neg\bigvee_{S \in \set{S}} \atj{\xth{\var{front\_S}}{T}}
.
\end{equation}
Trains that are leaving the network remain in the steady mode,
until they get away entirely\short{ and become idle}.

\newcommand{\TextAwayConditions}{%
Next, there are constraints related to the \var{idle} mode:
\begin{equation}\label{\eqL{phi:model:away:idle}}
( \atj{\xth{\var{enter}}{T}} \Rightarrow \neg\atj{\xth{\var{idle}}{T}} )
\ \land\
( \atj{\xth{\var{away}}{T}} \Rightarrow \atj{\xth{\var{idle}}{T}} )
.
\end{equation}
Then, there are some restrictions on variables in the next discrete step:
\begin{equation}\label{\eqL{phi:model:away:jump}}
( \atj{\xth{\var{enter}}{T}} \Rightarrow \neg\at{\xth{\var{enter}}{T}}{j+1} )
\ \land\
( \atj{\xth{\var{finished}}{T}} \Rightarrow \at{\xth{\var{finished}}{T}}{j+1} )
.
\end{equation}
Finally, restrictions of the related variables must hold---%
mutual exclusion of entering and finishing;
being away before entering the graph, and after finishing:
\begin{equation}\label{\eqL{phi:model:away:mutual}}
\begin{split}
\neg(\atj{\xth{\var{enter}}{T}} \land &\atj{\xth{\var{finished}}{T}} )
\ \land\
( \neg\atj{\xth{\var{away}}{T}} \Rightarrow \neg\at{\xth{\var{enter}}{T}}{j+1} )
\\\land\
\atj{\xth{\var{away}}{T}} \Rightarrow
    \ite[(][)]{&\atj{\xth{\var{finished}}{T}}}{\at{\xth{\var{away}}{T}}{j+1}}
        {\\\neg&\at{\xth{\var{finished}}{T}}{j+1} \land
            ( \at{\xth{\var{away}}{T}}{j+1} \lor \at{\xth{\var{enter}}{T}}{j+1} )}
.
\end{split}
\end{equation}
}

\full{\TextAwayConditions}

\paragraph{Transfer constraints}
control transferring of a~train to a next segment
when the end of one of the current segments is reached
(even when stopping).
We denote the fact that the back or front of train~$T$
reaches the end of segment \(S_1 \in \set{S}\)
by \atj{\xth{\var{pos}\var{\_exceed\_S}_1}{T}},
\(\var{pos} \in \{\var{back},\var{front}\}\),
which allows the train to move into\full{ segment}~$S_2$:
\begin{equation}\label{\eqL{phi:model:transfer}}
\begin{split}
\neg\atj{\xth{\var{idle}}{T}} \Rightarrow
    \bigwedge_{S_2 \in \set{S}, S_1 \rightarrow_T S_2} \bigl(
        (\atj{\xth{\var{pos}_1\var{\_S}_1}{T}} \land \atj{\xth{\var{pos}_2\var{\_S}_2}{T}})
            \Rightarrow&\\ \ite[(][)]
                {\atj{\xth{\var{pos}_1\var{\_exceed\_S}_1}{T}}}
                {\at{\xth{\var{pos}_1\var{\_S}_2}{T}}{j+1}&}
                {\at{\xth{\var{pos}_1\var{\_S}_1}{T}}{j+1}}
\bigr)
\end{split}
\end{equation}
where \(\var{pos}_1 \in \{\var{back},\var{front}\}\),
\(\var{pos}_2 = \Delta(\var{pos}_1)\),
\(\Delta = \{\var{back} \mapsto \var{front}, \var{front} \mapsto \var{next}\}\).
\newcommand{\TextTransferBack}{%
\short{%
\refEq{phi:model:transfer} defined regular train transfers within segments.
}%
For starting segments \(S \in \xth{\var{Start}}{T}\),
it is also necessary to eventually move
the back of the train inside the network
(which is not initially there, as stated in \refEq{phi:model:enter}):
\begin{equation}\label{\eqL{phi:model:transfer:start}}
\begin{split}
\Bigl(\neg\atj{\xth{\var{idle}}{T}} \land
    \neg\bigvee_{S \in \set{S}} \atj{\xth{\var{back\_S}}{T}}
    \Bigr) \Rightarrow
    \Bigl( \atj{\xth{\var{back\_inside}}{T}} \Leftrightarrow
        \bigvee_{S \in \xth{\var{Start}}{T}}
        \at{\xth{\var{back\_S}}{T}}{j+1} \Bigr)
,
\end{split}
\end{equation}
where \atj{\xth{\var{back\_inside}}{T}} means that
the train reaches the beginning of a~starting segment with its back.

In \var{idle} mode, no transfers happen from any segment~$S$:
\begin{equation}\label{\eqL{phi:model:transfer:idle}}
\Bigl( \atj{\xth{\var{idle}}{T}} \land \neg\at{\xth{\var{enter}}{T}}{j+1} \Bigr)
\Rightarrow \bigwedge_{\var{pos} \in \{\var{back}, \var{front}\}} \Bigl(
    \atj{\xth{\var{pos\_S}}{T}} \Leftrightarrow \at{\xth{\var{pos\_S}}{T}}{j+1} \Bigr)
.
\end{equation}
Also, when a~train is inside a~single segment, the back stays within:
\begin{equation}\label{\eqL{phi:model:transfer:stay}}
\bigl(\atj{\xth{\var{back\_S}}{T}} \land \atj{\xth{\var{front\_S}}{T}}\bigr)
    \Rightarrow \at{\xth{\var{back\_S}}{T}}{j+1}
.
\end{equation}
}%
\full{\TextTransferBack}

\full[When]{Finally, when}
a~train exceeds a~segment \(S \in \set{S}\) that is boundary,
the train is claimed as finished based on the front of the train:
\begin{equation}\label{\eqL{phi:model:transfer:finish}}
\atj{\xth{\var{front\_S}}{T}} \Rightarrow \ite[(][)]
    {\atj{\xth{\var{front\_exceed\_S}}{T}}}
    {\at{\xth{\var{finished}}{T}}{j+1}}
    {\at{\xth{\var{front\_S}}{T}}{j+1}}
,
\end{equation}
and it is claimed as away based on its back:
\begin{equation}\label{\eqL{phi:model:transfer:away}}
\atj{\xth{\var{back\_S}}{T}} \Rightarrow \ite[(][)]
    {\atj{\xth{\var{back\_exceed\_S}}{T}}}
    {\at{\xth{\var{away}}{T}}{j+1}}
    {\at{\xth{\var{back\_S}}{T}}{j+1}}
.
\end{equation}

\Img{consecutive}
    {A~conflicting plan of two consecutive trains with no stops}
    {\IncludeImg{0.9}{rail_consecutive}}

\paragraph{Mutual exclusion conditions}
prevent trains from collisions.
For each train~$T_1$ and for all segments~$S$,
all the mutual exclusion conditions are jointly
defined as
\begin{equation}\label{\eqL{phi:model:mutual}}
\bigwedge_{\var{pos}_1,\var{pos}_2 \in \set{P}}\
\bigwedge_{T_2 \in \set{T}, T_2 \neq T_1}
    \neg\bigl(
        \atj{\xth{\var{pos}_1\var{\_S}}{T_1}} \land
            \atj{\xth{\var{pos}_2\var{\_S}}{T_2}} \bigr)
.
\end{equation}
Thus, we require the segments adjacent to the current front segment to be free
(because \(\var{next} \in \set{P}\))---%
while it is whole sections\fullfootnote{%
In~\cite{rail-art}, such sections are called ``elementary routes''.
}
in the case of~\cite{rail-art},
as a~consequence of signal interlocking.
As a~result, tighter plans are possible in our case,
but the algorithm may also be forced to resolve
more violations of mutual exclusion conditions.
\refImg{consecutive} illustrates a~situation
where train~$A$ is followed by train~$B$
that enters as soon as train~$A$ leaves node~2.
Since the segment~2--3 is long,
train~$B$ will reach node~1 sooner than train~$A$ leaves node~3,
resulting in a~conflict at segment~2--3
that is claimed by train~$B$ as the next segment.

%%%%%%%%%%%%%%

\subsubsection{Initial Conditions.}\label{sss:phi:model:init}
At the beginning, each train stands still, either is away or starts its journey,
and is not finished.
And some train has to enter:
\begin{equation}\label{\eqL{phi:model:init}}
\begin{split}
&\bigwedge_{T \in \set{T}}\bigl(
\var{init}(\at{\xth{\fun{v}}{T}}{0}) = 0
\ \land\
( \at{\xth{\var{enter}}{T}}{0} \lor \at{\xth{\var{away}}{T}}{0} )
\ \land\
\neg\at{\xth{\var{finished}}{T}}{0}
\bigr)
\\\ \land\
&\bigvee_{T \in \set{T}} \at{\xth{\var{enter}}{T}}{0}
.
\end{split}
\end{equation}

%%%%%%%%%%%%%%

\subsubsection{Final Conditions.}\label{sss:phi:model:final}
In order to satisfy the connection constraints of trains completely,
we require the trains to
have finished moving through the network at the final unrolling step~$J$:
\begin{equation}\label{\eqL{phi:finish}}
\bigwedge_{T \in \set{T}}(
    \at{\xth{\var{finished}}{T}}{J} \land \at{\xth{\var{away}}{T}}{J}
)
.
\end{equation}

%%%%%%%%%%%%%%%%%%%%%%%%%%%%

\subsection{Schedule Constraints}\label{ss:phi:sched}
\full{%
Constraints of the model itself must be satisfiable,
otherwise it likely means that it is corrupt
(e.g. the infrastructure is disconnected).
Unsatisfiability of connection constraints is possible, but rarely desirable
(i.e. cases with connection lists where
it is not possible to visit all the nodes in the given order
wrt. the graph).
It is the schedule constraints that make the satisfiability problem interesting.
}

Schedule formulas enforce schedule constraints
and their Boolean combinations.
Orderings and timings described in \refSec{sss:problem:const:sched}
are translated into particular constraints
related to visiting nodes at discrete steps.
To encode such a~visit related to train \(T \in \set{T}\),
node \(N \in \set{N}\),
where~$\set{N}$ represents the set of nodes of the network,
and discrete step~$j$,
we use auxiliary Boolean variables \atj{\xth{\var{visit\_N}}{T}},
\(\var{visit} \in \{\var{arrive},\var{depart}\}\),
defined s.t.
\begin{equation}\label{\eqL{phi:sched:visit}}
\begin{split}
\atj{\xth{\var{arrive\_N}}{T}} &\Leftrightarrow
\left\{\begin{array}{ll}
    \bot, & \text{if } j = 0, \text{else}
\\  \bigvee_{S, S \rightarrow_T N}
        \at{\xth{\var{front\_S}}{T}}{j-1}
\\\qquad \land\ \atj{\xth{\var{finished}}{T}}
, & \text{if } N \in \xth{\var{End}}{T}, \text{ otherwise}
\\  \bigvee_{S, N \rightarrow_T S} \Big(
    \neg \at{\xth{\var{front\_S}}{T}}{j-1}
&       \land\ \atj{\xth{\var{front\_S}}{T}}\Big)
\\\qquad \land\ \neg\atj{\xth{\var{enter}}{T}}
;
\end{array}\right.\\
\atj{\xth{\var{depart\_N}}{T}} &\Leftrightarrow
\left\{\begin{array}{ll}
    \atj{\xth{\var{enter}}{T}}
    , & \text{\hspace{-2ex}if } N = \xth{\var{Start}}{T}, \text{else}
\\ \bot, & \text{\hspace{-2ex}if } j = 0 \lor N \in \xth{\var{End}}{T}, \text{ otherwise}
\\ \bigvee_{S, N \rightarrow_T S}
        \atj{\xth{\var{front\_S}}{T}}
&       \land\ \atj{\xth{\var{acc}}{T}}
        \land \var{init}(\atj{\xth{v}{T}}) = 0
,
\end{array}\right.
\end{split}
\end{equation}
where \(N \rightarrow_T S\) and \(S \rightarrow_T N\) means incidence
of the node~$N$ and segment \(S \in \set{S}\)
within the train~$T$'s connection,
in the corresponding direction;
%% In impl., there is always only one start/end, and only enter/finished is used
\xth{\var{Start}}{T} is the starting node of train~$T$,
and \xth{\var{End}}{T} is the set of the train's ending nodes.

%%%%%%%%%%%%%%

\subsubsection{Ordering.}\label{sss:phi:sched:order}
\refEq{problem:const:sched:order}
enforces \full[%
$\var{visit}_1$ to happen before $\var{visit}_2$ (\(\circ \in \{<, \leq \}\)).
This
]{%
an~order of $\var{visit}_1$ and $\var{visit}_2$.
Cases with \(\circ \in \{<, \leq \}\)
}%
requires to forbid $\var{visit}_2$ to take place before $\var{visit}_1$,
\emph{and} to make sure that $\var{visit}_2$ implies
that $\var{visit}_1$ already happened:
\begin{equation}\label{\eqL{phi:sched:order:<}}
\begin{split}
\bigwedge_{k=0}^J \Bigl( \at{\xth{\var{visit}_1\var{\_N}_1}{T_1}}{k} \Rightarrow
    \bigwedge_{l=0}^{K(k)} \neg \at{\xth{\var{visit}_2\var{\_N}_2}{T_2}}{l}
\Bigr)
\\\land
\bigwedge_{l=0}^J \Bigl( \at{\xth{\var{visit}_2\var{\_N}_2}{T_2}}{l} \Rightarrow
    \bigvee_{k=0}^{L(l)} \at{\xth{\var{visit}_1\var{\_N}_1}{T_1}}{k}
\Bigr)
,
\end{split}
\end{equation}
where \(K(k) = k\), \(L(l) = l-1\) if~$\circ$ is~$<$,
and \(K(k) = k-1\), \(L(l) = l\) if $\circ$ is~$\leq$.
\full{%
Orderings with~$\circ$ being~$>$ or~$\geq$ are simply handled
as cases with~$<$ or~$\leq$, respectively, with swapped arguments.

For cases where $\circ$ is~$=$, the produced formula is
\begin{equation}\label{\eqL{phi:sched:order:=}}
\bigwedge_{k=0}^J \Bigl( \at{\xth{\var{visit}_1\var{\_N}_1}{T_1}}{k} \Leftrightarrow
    \at{\xth{\var{visit}_2\var{\_N}_2}{T_2}}{k}
\Bigr)
.
\end{equation}
}

%%%%%%%%%%%%%%

\subsubsection{\short{Relative }Timing.}\label{sss:phi:sched:time}
In the first place, it is necessary to guarantee that the corresponding time condition holds
in cases when all the corresponding visits are active.
In cases where \(\circ \in \{<, \leq\}\), similarly to orderings,
we also make sure that violation of the timing implies
that the corresponding visits did already happen.
\full{\par%
Following \refEq{problem:const:sched:time:abs},
an~\emph{absolute timing} is translated into
\begin{equation}\label{\eqL{phi:sched:time:abs}}
\bigwedge_{k=0}^J \bigl( \at{\xth{\var{visit\_N}}{T}}{k} \Rightarrow
    \at{t}{k} \circ \xi \bigr)
\land \psi
.
\end{equation}
If \(\circ \in \{>, \geq\}\), then \(\psi \Leftrightarrow \top\),
otherwise
\begin{equation}\label{\eqL{phi:sched:time:abs:<}}
\psi \Leftrightarrow
\bigwedge_{k=0}^J \Bigl( \neg(\at{t}{k} \circ \xi) \Rightarrow
    \bigvee_{l=0}^{k-1} \at{\xth{\var{visit\_N}}{T}}{l}
\Bigr)
.
\end{equation}
\emph{Relative timing} constrains a~pair of visits.
}%
So \refEq{problem:const:sched:time:rel}
translates to
\begin{equation}\label{\eqL{phi:sched:time:rel}}
\bigwedge_{j=0}^J \Bigl( \at{\xth{\var{visit}_1\var{\_N}_1}{T_1}}{j} \Rightarrow
    \Bigl( \psi_j\ \land\
    \bigwedge_{k=j}^J \bigl( \at{\xth{\var{visit}_2\var{\_N}_2}{T_2}}{k}
    \Rightarrow (\at{t}{k} - \atj{t}) \circ \xi \bigr)
\Bigr)\Bigr)
\full[;]{,}
\end{equation}
\full{%
where
}%
\begin{equation}\label{\eqL{phi:sched:time:rel:<}}
\psi_j \Leftrightarrow
\left\{\begin{array}{ll}
    \top,& \text{ if } \circ \in \{>, \geq\};
\\  \bigwedge_{k=j}^J \Bigl(
        \neg\bigl((\at{t}{k} - \atj{t}) \circ \xi\bigr) \Rightarrow
            \bigvee_{l=j}^{k-1} \at{\xth{\var{visit}_2\var{\_N}_2}{T_2}}{l}
    \Bigr),& \text{ if } \circ \in \{<, \leq\}.
\end{array}\right.
\end{equation}

Since timings support
both lower and upper bounds,
and since Boolean combinations are allowed,
it is possible to~define interval boundaries, and more.

Recall that the variables~\atj{\var{t}}
only depend on the lengths~\atj{\var{\tau}} of integrations,
so the timing constraints are checked at the end of each integration.

%%%%%%%%%%%%%%%%%%%%%%%%%%%%

\newcommand{\SecFormulaSize}[1]{%
\subsection{Formula Size}\label{#1:phi:size}
Let $N_N$ be the size of the set of nodes~$\set{N}$
and $N_S$ the size of the set of segments~$\set{S}$.
Although \(N_S = \set{O}({N_N}^2)\),
the graphs that model railway networks are usually sparse,
thus \(N_S = \Theta(N_N)\).
The number of trains (i.e. size of~$\set{T}$)
should be negligible compared to~$N_N$ and~$N_S$.
For each train, the number of all the variables scales to~$\set{O}(J)$
(recall that $J$ is the last discrete step),
with the exception of auxiliary variables \atj{\xth{\var{visit\_N}}{T}},
which scale to $\set{O}(J \cdot N_N)$,
and the variables \atj{\xth{\var{pos\_S}}{T}},
which scale to $\set{O}(J \cdot N_S)$.

The number of particular differential equations and invariants, for each train,
scales to $\set{O}(J)$ as well.
The number of other atomic predicates, that are part of the railway model,
related to a~separate train or not,
scales to $\set{O}(J \cdot {N_S}^2)$.

In the case of schedule constraints,
especially \refEq{phi:sched:order:<} and \refEq{phi:sched:time:rel},
the number of produced atomic predicates
corresponds to a~polynomial of~$J$ (assuming that the graph is sparse):
$\set{O}(J^3)$ in the case of relative timings
with the operator~$<$ or~$\leq$
(\refEq{phi:sched:time:rel} and \refEq{phi:sched:time:rel:<}),
and $\set{O}(J^2)$ in the rest cases.
}

\full{\SecFormulaSize{ss}}

%%%%%%%%%%%%%%%%%%%%%%%%%%%%

\newcommand{\SecInputLanguage}[1]{%
\subsection{Input Language}\label{#1:phi:lang}
Input formula format is derived from SMT-LIB~\cite{smtlib-reference-art},
with a~syntax similar to QF\_\-NRA logic
(quantifier-free non-linear theory of reals),
but wrt. features described in \refSec{s:theory}.
In particular, the language is extended of description of \acr{ode} systems.
A~specification of the core language
is available~\opencite{unsot-core_lang}.
Also, to~support forming formulas in a~generic way,
we developed a~preprocessing language~\opencite{unsot-preprocess_lang}
that is actually embedded right within an~input of the solver,
and is usable for other SMT-LIB logics too.
As a~result, the whole railway-specific encoding
is provided in the form of a~library of the preprocessing language.
}

\full{\SecInputLanguage{ss}}

%%%%%%%%%%%%%%%%%%%%%%%%%%%%%%%%%%%%%%%%%%%%%%%%%%%%%%%%%

\section{Algorithm}\label{s:alg}
We solve the benchmark problem using an improvement of an SMT solver
\open[that was]{we} introduced earlier~\cite{my-tap-art}.
The solver is based on \acr{dpll} with \acrf{cdcl},
but improves the original naive
lazy offline approach to a~lazy online approach with exhaustive theory
propagation~\cite{smt-dpll_t}
to support efficient handling of both
Boolean and theory constraints.
\full[%
The rest of this section requires basic knowledge
of \acr{sat} and \acr{smt} solving{\open[
(see \cite{smt-dpll_t} or \refAppend{a:alg})%
]{~\cite{smt-dpll_t,my-rail-online-art}%
}}.]{%
The online approach is quite a~common technique of nowadays \acr{smt} solvers.
This uses a~modified version of the SAT solver with callbacks
inserted into the most important parts of a~DPLL algorithm.
}

\newcommand{\ParDpllCdcl}{%
\paragraph{DPLL with CDCL.}
Vast of nowadays SAT solvers are based on \acrf{dpll} algorithm,
which introduced basic techniques of the search procedure:
\emph{unit propagation},
which forces an~assignment that avoids a~conflict
(i.e., an~assignment of a~set of variables that would cause unsatisfiability);
\emph{decide}, which guesses a~value for a~variable (usually based on some heuristics)
that is not assigned yet,
and usually only in cases
when no other technique can be applied at the moment (especially propagations);
and \emph{backtrack}, which reverts the recent decision when resolving a~conflict.

Another technique, that is implemented in most state-of-the-art SAT solvers,
is called \acrf{cdcl}.
After arriving at a~conflict, it remembers the incompatible decisions,
that caused the conflict,
in the form of a~conflict clause, which is inserted into the clause set.
Minimal size of a~conflict clause can, as a~result, prune the searched state space significantly.
Conflict clauses are often related to an~enhancement of the backtracks,
called \emph{backjumping},
which returns directly back to a~decision that caused the conflict,
instead of reverting all the decisions that are not necessarily related to the conflict.

Such an~approach always terminates (e.g., no deadlock is possible)~\cite{smt-dpll_t},
and the discovered conflicts are efficiently resolved.

\medskip

The rest of the section describes techniques that are related to \acr{smt} solving.
}

\full{\ParDpllCdcl}

\newcommand{\ParLazyApproach}{%
\paragraph{Lazy approach}
is based on lazy evaluation,
which means that theory-specific constraints
are used only when they are needed.
The procedure is based on cooperation
of a~SAT solver and a~theory solver, so-called \set{T}-solver.

The SAT solver computes only with \emph{abstracted} models,
meaning that theory-related predicates are abstracted to pure Boolean variables.
Unsatisfiability of the abstracted model implies
unsatisfiability of the original formula,
regardless of the underlying theory,
and satisfiability implies
that it \emph{can} be also \emph{consistent} with the theory,
which is when the \set{T}-solver kicks in.
If it is considered inconsistent,
some explanation of the inconsistency is added
into the propositional constraints set,
in the form of a~conflict clause (as expected for a~\acr{cdcl} algorithm),
and other Boolean model is sought.
This process is repeated until the SAT solver
finds a~consistent model, or returns unsatisfiable.

While the SAT solver is guaranteed not to,
\set{T}-solver can actually block, depending on the underlying theory
(e.g., it is possible in the case of~\cite{rail-art}).
In our theory, it may happen that an~integration never ends,
when all the invariants result in tautology, which stems from concrete model.
This is not the case of the presented railway scheduling model, though,
so no deadlock is possible there.
}

\full{\ParLazyApproach}

Our theory solver is based on the floating-point simulation semantics
of the theory described in \refSec{s:theory}.
It uses equalities with only a~single variable on one of their two sides
and differential constraints
as \emph{inference rules}~\cite{my-tap-art}
that may assign values to the corresponding isolated variables.
For example, if the values of~\at{t}{j-1} and~\at{\tau}{j-1}
in \(\atj{t} = \at{t}{j-1} + \at{\tau}{j-1}\) are already fixed,
then we can infer the value of \atj{t}.
In a~similar way,
if the initial value of \atj{\xth{\fun{v}}{T}}
and the value of \atj{\xth{a}{T}} is fixed,
then we can use \(\atj{\xth{\der{v}}{T}} = \atj{\xth{a}{T}}\)
to infer the value of the functional variable \atj{\xth{\fun{v}}{T}}.
All other constraints are numerically evaluated
as soon as all their variables have assigned values.
This is not complete in general,
but suffices to eventually decide the problems
necessary for solving the planning problems described in this article.

Atomic predicates form vertices of a~directed \emph{dependency graph},
where an~edge means that
the source vertex is an~inference rule that may assign a~value to
a~floating-point variable that is shared with the target vertex.
Inference rules corresponding to vertices
with no input edges are \emph{initial} inference rules.

\newcommand{\ParOnlineApproach}{%
\paragraph{Online approach}
requires an~underlying SAT solver to~support
\emph{partial} propositional assignments
and to~allow the \set{T}-solver to~interfere
the procedure via callbacks.
If, for example, consistency is being checked
after setting each Boolean variable,
then an~inconsistency can be found early.
In any way, the \set{T}-solver must keep track with the SAT solver
and the currently assigned Boolean variables,
and also wrt. to the theory model and/or assignments of variables
(e.g., floating-point variables).
Online approach also allows \emph{backjumping},
instead of restarting from scratch.
}

\full{\ParOnlineApproach}

\newcommand{\ParTheoryPropagation}{%
\paragraph{Theory propagation}
is maybe the most important technique of our lazy approach.
The algorithm does not only check the consistency, but actually guides the search
according to the theory axioms.
In other words,
when an~abstracted theory predicate is assigned, all other predicates
that relate to the assigned one, according to the theory,
are forced to a~proper value, if it is clear according to the axioms.
For example, if the SAT solver assigns
a~predicate \(x = 0\) to~$\top$ in our theory,
and another, still unassigned predicate \(x = 1\) is present
in the constraints set, then the \(x = 1\) is forced to~$\bot$.
This prevents a~lot of inconsistencies from happening.
Without this, the SAT solver would have to do only unit propagations,
or worse, a~decision whether to~assign \(x = 1\) to~$\bot$ or~$\top$,
because it would have no clue which one to~choose\footnote{%
In this particular case, a~preprocessing like \(x = 0 \Rightarrow \neg(x = 1)\)
would help, but such exclusions are generally difficult to~resolve
within the preprocessing stage,
if the right hand side of the equations are more complicated.
}.

If checking the consistency of the current assignment
and theory propagation is a~cheap process,
it can be applied literally after any Boolean assignment
of a~theory atomic predicate.
In terms of theory propagation,
such a~strategy is called \emph{exhaustive theory propagation},
where, actually, the consistency is usually being checked exhaustively too.
}

\full[\paragraph{Theory propagation.}
We perform exhaustive theory propagation,
along with consistency checks,
]{%
\ParTheoryPropagation
This is our case,
}%
because all inference rules are based on floating-point arithmetic,
which is cheap\footnote{%
Simulations of ODEs are actually not that cheap,
but we currently do not have evidence that
postponing them within theory propagation would be beneficial.
%% Jeste jsem to nezkoumal ...
}.
Constraints that are currently not evaluable cannot be propagated nor checked for consistency,
though.

\full{%
It is sufficient to~propagate
only inference rules that assigned a~floating-point variable,
and to~traverse only predicates that \emph{depend} on the rule.
For example, \(p^>: x>0\) will only be checked for consistency,
but in the case of \(p^=: x=0\), \(p^= \leftarrow \top\)
will propagate $p^>$ to~$\bot$.

\hide{%
%%! It would not be theory-propagation, but rather a decision heuristic
%%! It is NOT clear if we can propagate such a predicate to TRUE
The dependencies of theory constraints are transitive, starting from initial values.
To~prevent theory propagations from being too eager,
we do not force inference rules to~$\top$ if such an~enforcement would cause an~assignment.
In such cases, propagations of the SAT solver are preferred.
For example, for \(p^-: y=x-1\) and \(p^+: y=x+1\),
and for \(\ite{p^>}{p^-}{p^+}\),
\(p^= \leftarrow \top\) will again theory-propagate $p^>$ to~$\bot$,
but will not propagate $p^-$ nor~$p^+$ (where the order would matter).
Instead, consequent unit propagation of~$p^>$ will correctly cause
\(p^+ \leftarrow \top\),
which will theory-propagate $p^-$ to~$\bot$.
}
}

\paragraph{Decision heuristics.}
In the case of formulas with a~structure similar to a~BMC unrolling,
each consecutive step
depends on the values from the previous one.
Thus, a~suitable strategy, called BMC strategy, is to first decide Booleans
that correspond to the lower steps.

In our case, it is often useful to prefer deciding inference rules that can be used,
for example, initial inference rules, or those that depend on the already evaluated ones,
based on the dependency graph.
Then, the inference rule allows theory propagation,
which may then enable consistency checks.
Thus, within the same discrete step, we prefer
initial inference rules,
then the other inference rules,
then other predicates, and lastly pure Booleans.

\full{%
We list also another possible strategy---%
to ultimately prefer the ``most initial'' inference rules.
To~achieve that, we apply a~modified version of the Floyd–Warshall algorithm
within the preprocessing stage, where we are interested
in distances between predicates within the dependency graph.
For example, for \(p_1: x_0=0\), \(p_2: x_1=x_0 + y\) and \(p_3: x_1 < 0\),
the distance from~$p_1$ to~$p_3$ is~2.
Then, floating-point variables are sorted
s.t. the ones with a~most distant predicate
from a~corresponding inference rule comes first.
For example, $x_0$ comes before~$x_1$.
Following such an~order, a~predicate of one of the corresponding variables
is decided to~$\top$,
preferring inference rules over other predicates.
This way, the conflict clauses resulting from the discovered inconsistencies
are short, because the decisions are being made not far from initial conditions
(i.e. initial inference rules).
Also, the resulting order often corresponds
to the actual order of BMC-like unrollings,
even though this information is not utilized explicitly.
\full{%
The order of BMC discrete steps is not followed in the case
of constraints on the last discrete step,
because such constraints
actually may meet the definition of initial inference rules.
}%
\full[A~]{%
Note that the computation time of all the necessary distances
in the dependency graph
is not always negligible.
Another
}%
drawback is
that the strategy does not consider pure Boolean variables at all.
These can be important decision variables,
like \atj{\xth{\var{enter}}{T}} or \atj{\xth{\var{next\_S}}{T}}
in the case of the presented railway scheduling problem.
As a~result, some predicates
(e.g. \(\atj{\xth{a}{T}} = \atj{\xth{A}{T}}\)), that could be propagated
based on such Boolean decisions,
are being decided instead,
because such predicates have higher decision priority than pure Booleans.
\full{%
This can lead to unnecessary inconsistencies,
which could be prevented by unit propagations and theory propagations.
}}%
\full{%
In our experiments, this heuristic was less efficient than the BMC strategy
on average, but not too significantly, and not in all test cases.
}

In the case of railway scheduling, we designed
a~strategy that is specific to the given task.
We modified the BMC strategy s.t. the Booleans
\atj{\xth{\var{enter}}{T}}, \atj{\xth{\var{idle}}{T}},
and \atj{\xth{\var{next\_S}}{T}}
are additionally set to the highest decision priority within each step~$j$,
in the listed order.
Here we first set
\atj{\xth{\var{enter}}{T}} to~$\top$
and \atj{\xth{\var{idle}}{T}} to~$\bot$
to~prefer that the trains finish as soon as possible. We first set
\atj{\xth{\var{next\_S}}{T}} to~$\bot$
to~avoid activation of a segment before being unit-propagated using
\refEq{phi:model:pos:next}.
\full{%
Again, the results are better than for the previous strategies on average,
but not too significantly.
}

\full{%
The SAT solver has its own decision heuristic too.
We use it as a~fallback strategy,
when no other decision rule is available,
for example in the case of pure Booleans in the ``initial'' strategy.
The heuristic can be designed to, for example,
prefer variables that frequently participate in Boolean conflicts.
Such a~strategy is beneficial, because the ``problematic variables''
are being resolved soon.
}

\full{%
So far, we discussed only ``static'' strategies,
which are precomputed within the preprocessing stage
and does not adapt to a~current assignment of variables
(except of the mentioned heuristic of the SAT solver).
It is expectable that such sophisticated strategies
can do even better, especially with a~knowledge of an~underlying task,
and with utilizing machine learning techniques.
We have not implemented such a~``dynamic'' strategy yet.
}

%%%%%%%%%%%%%%%%%%%%%%%%%%%%%%%%%%%%%%%%%%%%%%%%%%%%%%%%%

\section{Experimental Part}\label{s:exp}
In Sections~\ref{s:intro},~\ref{s:problem} and~\ref{s:phi},
we mentioned differences of our model and algorithm
compared to an~approach that is based on dedicated railway simulations~\cite{rail-art}.
Although we support a~richer set of schedule constraints,
here we stick to case studies that can be handled by both approaches.
\short[Our ]{%
Still, we omit numerical comparisons (e.g. the absolute run-times)
here, since the respective tools solve different problems
and a~thorough discussion is needed---%
details can be found
{\open[in \refAppend{a:exp}]{in the extended version of the paper~\cite{my-rail-online-art}}.}
Especially, our
}%
model
is not based on signal interlocking
and exhibits more nondeterminism
(e.g., we allow the trains to wait in stations and before entering the network).
\full[Hence, we ]{%
There are more differences between both the approaches, though.
We firstly
}%
focus on a~qualitative analysis of the behavior of the tools\full{,
where the differences are not that significant.
Then, we also show numerical comparisons (e.g. the absolute run-times)
of the tools, with a proper discussion%
}.

We use our model from \refSec{s:phi} and our implementation~\opencite{unsot}
of the algorithm from \refSec{s:alg},
and the \emph{rail\-perf\-check} tool~\cite{rail-art}.
We focus on case studies where it is not trivial to decide whether a~plan
that meets both ordering and timing constraints exists\full{,
that is, if the formula is satisfiable (\Sat{}), or not (\Unsat{})%
}.
For this, we generalized the experiments named \emph{Gen} in~\cite{rail-art},
where all the other experiments, in contrast, exhibit easily satisfiable schedule constraints,
which should not be challenging for approaches that are based on SAT solving.
\hide{%
We also present some related cases
that are not possible to~handle with \emph{rail\-perf\-check},
so without a~comparison.
}

\newcommand{\TextEquipment}{%
Both presented tools use Minisat~\cite{minisat} as the underlying SAT solver.
We use Odeint~\cite{odeint-art} as the ODE solver.
We exclude execution time of our preprocessing (i.e., generating the formula),
which we did not optimize\footnote{%
In the worst case (\(N_T = N_S = 4\),
scenario \emph{all}), the preprocessing takes 12~minutes.
}.
}

\full{\TextEquipment}

\paragraph{Specification.}
\Img{serial-parallel}
    {An~example of a~serial-parallel infrastructure, with $N_S = 2$ and $N_P = 3$}
    {\IncludeImg{0.85}{rail_serial_parallel}}

We use a~serial-parallel network for our experiments---%
a~track with $N_S$ serially connected
groups of $N_P$~identical parallel tracks with a~station.
See an~example in \refImg{serial-parallel}.
For our experiments, we will assume \(N_S = N_P\).
We use \full{equivalent }trains \(\set{T} = \{T_1, \dots, T_{N_T}\}\)
with acceleration rate \(A=2\), deceleration rate \(B=1\),
velocity limit~\(\var{V_{max}}=40\) and length \(L=50\).
Each train is assigned to connection list $(\var{start}, \var{end})$,
which only contains the boundary nodes.
As a~result, multiple paths are possible ($N_P^{N_S}$, at most) for each train.
Also, the trains are not allowed to stop at any station,
but just drive through,
once they enter the network.
In the case of~\cite{rail-art},
it is not possible to force the trains not to stop at stations
\emph{and} to make them drive consecutively after each other,
due to signal interlocking\short{%
{\open[(see the appendix)]{~\cite{my-rail-online-art}}}%
}.
However, this fact \full[%
does not affect the analysis provided in this section%
]{%
affects only the numeric comparisons,
not the qualitative analysis%
}.

\newcommand{\TextSpecNop}{%
\full[We also]{First, we} define a~scenario \var{nop}\short{,
equivalent to the experiments in \refSec{s:exp}, but
}%
with no schedule constraints at all,
to~show that the trains can finish within the given $J$~discrete steps---%
if~$J$ was too low, the result could generally be \Unsat{},
even with no violations of schedule constraints.
}%
\full{%
\TextSpecNop
}%
In our case,
we selected the number of unrollings~$J$ manually for each particular experiment---%
high enough
to~allow all the trains to~finish (i.e., to~satisfy \refEq{phi:finish}).
Such a~parameter is not needed in the case of \emph{rail\-perf\-check}.

\full[We present two]{Next, there are two regular}
scenarios, \var{last} and \var{all},
that are defined as follows:
\begin{itemize}
\item \var{last}: the last train~$T_{N_T}$ must satisfy a~relative timing,
    and the other trains~$T_i$ just enter in a~given order:
    \begin{equation}\label{\eqL{exp:sched:last}}
    \begin{split}
    \var{timing}(T_{N_T}, \var{bnd})\hide{\var{timing}(T_{N_T}, \circ, \var{bnd})} \land
    \bigwedge_{i < N_T} \bigl(
        \var{enter_{before}}(T_i, T_{i+1}) \land
        \var{early_{after}}(T_{i+1}, T_i)
    \bigr)
    ,
    \end{split}
    \end{equation}
\item \var{all}: each particular train~$T_i$ must satisfy a~relative timing:
    \begin{equation}\label{\eqL{exp:sched:all}}
    \begin{split}
    \bigwedge_i \Bigl(&
    \var{timing}(T_i, \var{bnd})\hide{\var{timing}(T_i, \circ, \var{bnd})}
    \ \land
    \\
    \Bigl(&\var{enter_{first}}(T_i) \ \lor
        \bigvee_{j \neq i} \bigl(
            \var{enter_{before}}(T_j, T_i) \land
            \var{early_{after}}(T_i, T_j)
    \bigr) \Bigr)
    \Bigr)
    ,
    \end{split}
    \end{equation}
\end{itemize}
where \(T_i,T_j \in \set{T}\), and
with
\begin{itemize}
\item \(\var{timing}(T, \var{bnd})\hide{\var{timing}(T, \circ, \var{bnd})} \Leftrightarrow
    \var{transfer}(\var{departure}(T, \var{start}),
        \var{arrival}(T, \var{end})) \full[<]{\circ} \var{bnd}\),
\item \(\var{enter_{before}}(T_1, T_2) \Leftrightarrow
    \var{departure}(T_1, \var{start}) < \var{departure}(T_2, \var{start})\),
\item \(\var{early_{after}}(T_1, T_2) \Leftrightarrow
    \var{departure}(T_1, \var{start}) \leq \var{arrival}(T_2, \var{end}_1)\), and
\item \(\var{enter_{first}}(T) \Leftrightarrow
    \var{departure}(T, \var{start}) = 0\),
\end{itemize}
where $\var{end}_1$ is the joint node $E_1$ in the figure.
The purpose of \var{early_{after}} along with \var{enter_{before}}
is to avoid long gaps between two consecutive trains,
to~reduce the amount of nondeterminism of waiting of the trains.
\full{%
Parallel segments which come into the node $\var{end}_1$
are the first ones which do not block a~preceding train from entering
(according to \refEq{phi:model:mutual}).
As a~result, a~next train is allowed to enter
only as long as the successor drives within these segments.
}%
In the case of \emph{rail\-perf\-check}, this is not necessary since
waiting is deterministic.
Note that in scenario \var{last}, trains are fully ordered,
while in scenario \var{all}, they are not ordered at all.

\hide{%
To~supply a~further insight, we provide
\var{last_{abs}} and \var{all_{abs}} scenarios,
in addition.
These modify the corresponding scenarios by using absolute instead of relative timing:
\(\full[\var{timing_{abs}}(T, \var{bnd})]{\var{timing_{abs}}(T, \circ, \var{bnd})} \Leftrightarrow
\var{arrival}(T, \var{end}) \full[<]{\circ} \var{bnd}\).
Such a~constraint cannot be achieved with \emph{rail\-perf\-check}.
}

Each case study is parametrized
by a~scenario, variables
\(N_T,N_S \in \{1,2,3,4\}\),
and a~timing upper\hide{ or lower} bound \(\var{bnd} \in \{10^1,10^2,10^3\}\).
In our case, additionally, \(\rho = 30\)
(timeout for functional variables in \refSec{ss:phi:model}),
and \(J = \Gamma(N_T)\),
with \(\Gamma = \{ 1 \mapsto 45, 2 \mapsto 80, 3 \mapsto 115, 4 \mapsto 150 \}\).
In the case of~\cite{rail-art},
$J$ is incrementally increased up to \(2 \cdot N_T\).
\full{%
Since timings with lower bounds and absolute timings
are not supported by \emph{rail\-perf\-check},
we omit them.
}

Both scenarios \emph{last} and \emph{all} are equivalent
in cases with only one train (\(N_T = 1\)).
These are the cases named \emph{Gen} in~\cite{rail-art}.

\paragraph{Results.}
\dashlinedash=1pt
\dashlinegap=2pt
\newcommand{\TableResultsNop}{%
\hide{%
\csvset{myStyle/.append style={
    table head=\hline\csvlinetotablerow\\\hline\hline,
}}}
\Table{exp:nop}{0.9}
    {Running time comparison of \var{nop} scenario}
    {\hide{\CsvTabular{nop}{rr|l||ll|l}}
    \footnotesize\begin{tabular}{|rr|l||ll|l|}
\hline
$N_T$ & $N_S$ & Result & Our & conflicts & Their \\\hline\hline
1 & 2 & \texttt{sat} & 0.1 s & 0 & 0 s \\
  & 3 & \texttt{sat} & 0.1 s & 0 & 0 s \\
  & 4 & \texttt{sat} & 0.1 s & 0 & 0 s \\\hline
2 & 2 & \texttt{sat} & 0.4 s & 4 & 0 s \\
  & 3 & \texttt{sat} & 0.6 s & 26 & 0.1 s \\
  & 4 & \texttt{sat} & 1 s & 51 & 0.7 s \\\hline
3 & 2 & \texttt{sat} & 4.3 s & 29 & 0 s \\
  & 3 & \texttt{sat} & 8.3 s & 284 & 0.3 s \\
  & 4 & \texttt{sat} & 17 s & 684 & 2.1 s \\\hline
4 & 2 & \texttt{sat} & 39 s & 106 & 0.1 s \\
  & 3 & \texttt{sat} & 1.5 m & 1559 & 0.6 s \\
  & 4 & \texttt{sat} & 4 m & 5100 & 4.3 s \\\hline
\end{tabular}
}
}
\newcommand{\TableResultsLt}{%
\hide{%
\csvset{myStyle/.append style={
    table head=\hline\csvlinetotablerow\\,
}}}
\Table{exp:lt}{0.9}
    {Running time comparison of \var{last} and \var{all} scenarios\hide{ with timing $<\var{bnd}$}}
    {\hide{\CsvTabular{lt_short}
    {rrr|l|l@{\hspace{-.5ex}}c@{\hspace{-.5ex}}l
           |l@{\hspace{0ex}}c@{\hspace{0ex}}l
    ||rrr|l|l@{\hspace{-.5ex}}c@{\hspace{-.5ex}}l
           |l@{\hspace{0ex}}c@{\hspace{0ex}}l
    }\hide{\CsvTabular{lt}
    {rrr|l||l@{\hspace{-.5ex}}c@{\hspace{-.5ex}}l
           |l@{\hspace{0ex}}c@{\hspace{0ex}}l
           |l|l
    }}}
    \footnotesize\begin{tabular}{%
|rrr|l|l@{\hspace{-.5ex}}c@{\hspace{-.5ex}}l
    |l@{\hspace{0ex}}c@{\hspace{0ex}}l
||rrr|l|l@{\hspace{-.5ex}}c@{\hspace{-.5ex}}l
    |l@{\hspace{0ex}}c@{\hspace{0ex}}l|}
\hline
 &  &  &  &  & $last$ &  &  & $all$ &  &  &  &  &  &  & $last$ &  &  & $all$ & {} \\
$N_T$ & $N_S$ & $bnd$ & Result & Our &  & Their & Our &  & Their & $N_T$ & $N_S$ & $bnd$ & Result & Our &  & Their & Our &  & Their \\
 &  &  &  &  &  &  &  &  &  &  &  &  &  &  &  &  &  &  & {} \\\hline\hline
1 & 2 & $10^1$ & \texttt{unsat} &  &  &  & 0.1 s &  & 0 s & 3 & 2 & $10^1$ & \texttt{unsat} & 11 s &  & 4.6 s & 0.6 s &  & 24 s \\
 &  & $10^2$ & \texttt{unsat} &  & $\rightarrow$ &  & 0.1 s &  & 0 s &  &  & $10^2$ & \texttt{unsat} & 1.1 m &  & 4.6 s & 2.6 m &  & 24 s \\
 &  & $10^3$ & \texttt{sat} &  &  &  & 0.1 s &  & 0 s &  &  & $10^3$ & \texttt{sat} & 3.7 s &  & 0 s & 4.2 s &  & 0 s \\\hdashline
 & 3 & $10^1$ & \texttt{unsat} &  &  &  & 0.1 s &  & 0.3 s &  & 3 & $10^1$ & \texttt{unsat} & 38 s &  & $>2$ h & 0.9 s &  & $>2$ h \\
 &  & $10^2$ & \texttt{unsat} &  & $\rightarrow$ &  & 0.1 s &  & 0.3 s &  &  & $10^2$ & \texttt{unsat} & 13 m &  & $>2$ h & 16 m &  & $>2$ h \\
 &  & $10^3$ & \texttt{sat} &  &  &  & 0.1 s &  & 0 s &  &  & $10^3$ & \texttt{sat} & 6.8 s &  & 0.3 s & 8.3 s &  & 0.3 s \\\hdashline
 & 4 & $10^1$ & \texttt{unsat} &  &  &  & 0.1 s &  & 3.5 s &  & 4 & $10^1$ & \texttt{unsat} & 2.2 m &  & $>2$ h & 1.1 s &  & $>2$ h \\
 &  & $10^2$ & \texttt{unsat} &  & $\rightarrow$ &  & 0.2 s &  & 3.5 s &  &  & $10^2$ & \texttt{unsat} & 1.1 h &  & $>2$ h & 1 h &  & $>2$ h \\
 &  & $10^3$ & \texttt{sat} &  &  &  & 0.1 s &  & 0 s &  &  & $10^3$ & \texttt{sat} & 13 s &  & 2.1 s & 16 s &  & 2.1 s \\\hline
2 & 2 & $10^1$ & \texttt{unsat} & 0.4 s &  & 0.6 s & 0.2 s &  & 0.9 s & 4 & 2 & $10^1$ & \texttt{unsat} & 3.6 m &  & 33 s & 1.2 s &  & 9.3 m \\
 &  & $10^2$ & \texttt{unsat} & 2.2 s &  & 0.6 s & 2.5 s &  & 0.9 s &  &  & $10^2$ & \texttt{unsat} & 13 m &  & 33 s & 21 m &  & 9.3 m \\
 &  & $10^3$ & \texttt{sat} & 0.4 s &  & 0 s & 0.4 s &  & 0 s &  &  & $10^3$ & \texttt{sat} & 25 s &  & 0.1 s & 39 s &  & 0.1 s \\\hdashline
 & 3 & $10^1$ & \texttt{unsat} & 0.8 s &  & 1.8 m & 0.3 s &  & 3.5 m &  & 3 & $10^1$ & \texttt{unsat} & 31 m &  & $>2$ h & 1.4 s &  & $>2$ h \\
 &  & $10^2$ & \texttt{unsat} & 8.9 s &  & 1.8 m & 9.2 s &  & 3.5 m &  &  & $10^2$ & \texttt{unsat} & $>2$ h &  & $>2$ h & $>2$ h &  & $>2$ h \\
 &  & $10^3$ & \texttt{sat} & 0.7 s &  & 0.1 s & 0.7 s &  & 0.1 s &  &  & $10^3$ & \texttt{sat} & 51 s &  & 0.6 s & 1.4 m &  & 0.6 s \\\hdashline
 & 4 & $10^1$ & \texttt{unsat} & 1.6 s &  & $>2$ h & 0.5 s &  & $>2$ h &  & 4 & $10^1$ & \texttt{unsat} & $>2$ h &  & $>2$ h & 2.1 s &  & $>2$ h \\
 &  & $10^2$ & \texttt{unsat} & 26 s &  & $>2$ h & 29 s &  & $>2$ h &  &  & $10^2$ & \texttt{unsat} & $>2$ h &  & $>2$ h & $>2$ h &  & $>2$ h \\
 &  & $10^3$ & \texttt{sat} & 1.1 s &  & 0.7 s & 1.1 s &  & 0.7 s &  &  & $10^3$ & \texttt{sat} & 2 m &  & 4.3 s & 3.9 m &  & 4.3 s \\\hline
\end{tabular}

    \smallskip

    According to $N_S$, total number of nodes is
    \(2 \mapsto 10, 3 \mapsto 17, 4 \mapsto 26\),
    and more importantly, the number of possible paths for each train is
    \(2 \mapsto 4, 3 \mapsto 27, 4 \mapsto 256\).
}}

\full{%
\hide{%
\csvset{myStyle/.append style={
    table head=\hline\csvlinetotablerow\\\hline\hline,
}}}
\Table{exp:nop}{0.9}
    {Running time comparison of \var{nop} scenario}
    {\hide{\CsvTabular{nop}{rr|l||ll|l}}
    \footnotesize\begin{tabular}{|rr|l||ll|l|}
\hline
$N_T$ & $N_S$ & Result & Our & conflicts & Their \\\hline\hline
1 & 2 & \texttt{sat} & 0.1 s & 0 & 0 s \\
  & 3 & \texttt{sat} & 0.1 s & 0 & 0 s \\
  & 4 & \texttt{sat} & 0.1 s & 0 & 0 s \\\hline
2 & 2 & \texttt{sat} & 0.4 s & 4 & 0 s \\
  & 3 & \texttt{sat} & 0.6 s & 26 & 0.1 s \\
  & 4 & \texttt{sat} & 1 s & 51 & 0.7 s \\\hline
3 & 2 & \texttt{sat} & 4.3 s & 29 & 0 s \\
  & 3 & \texttt{sat} & 8.3 s & 284 & 0.3 s \\
  & 4 & \texttt{sat} & 17 s & 684 & 2.1 s \\\hline
4 & 2 & \texttt{sat} & 39 s & 106 & 0.1 s \\
  & 3 & \texttt{sat} & 1.5 m & 1559 & 0.6 s \\
  & 4 & \texttt{sat} & 4 m & 5100 & 4.3 s \\\hline
\end{tabular}
}

\hide{%
\csvset{myStyle/.append style={
    table head=\hline\csvlinetotablerow\\,
}}}
\Table{exp:lt}{0.9}
    {Running time comparison of \var{last} and \var{all} scenarios\hide{ with timing $<\var{bnd}$}}
    {\hide{\CsvTabular{lt_short}
    {rrr|l|l@{\hspace{-.5ex}}c@{\hspace{-.5ex}}l
           |l@{\hspace{0ex}}c@{\hspace{0ex}}l
    ||rrr|l|l@{\hspace{-.5ex}}c@{\hspace{-.5ex}}l
           |l@{\hspace{0ex}}c@{\hspace{0ex}}l
    }\hide{\CsvTabular{lt}
    {rrr|l||l@{\hspace{-.5ex}}c@{\hspace{-.5ex}}l
           |l@{\hspace{0ex}}c@{\hspace{0ex}}l
           |l|l
    }}}
    \footnotesize\begin{tabular}{%
|rrr|l|l@{\hspace{-.5ex}}c@{\hspace{-.5ex}}l
    |l@{\hspace{0ex}}c@{\hspace{0ex}}l
||rrr|l|l@{\hspace{-.5ex}}c@{\hspace{-.5ex}}l
    |l@{\hspace{0ex}}c@{\hspace{0ex}}l|}
\hline
 &  &  &  &  & $last$ &  &  & $all$ &  &  &  &  &  &  & $last$ &  &  & $all$ & {} \\
$N_T$ & $N_S$ & $bnd$ & Result & Our &  & Their & Our &  & Their & $N_T$ & $N_S$ & $bnd$ & Result & Our &  & Their & Our &  & Their \\
 &  &  &  &  &  &  &  &  &  &  &  &  &  &  &  &  &  &  & {} \\\hline\hline
1 & 2 & $10^1$ & \texttt{unsat} &  &  &  & 0.1 s &  & 0 s & 3 & 2 & $10^1$ & \texttt{unsat} & 11 s &  & 4.6 s & 0.6 s &  & 24 s \\
 &  & $10^2$ & \texttt{unsat} &  & $\rightarrow$ &  & 0.1 s &  & 0 s &  &  & $10^2$ & \texttt{unsat} & 1.1 m &  & 4.6 s & 2.6 m &  & 24 s \\
 &  & $10^3$ & \texttt{sat} &  &  &  & 0.1 s &  & 0 s &  &  & $10^3$ & \texttt{sat} & 3.7 s &  & 0 s & 4.2 s &  & 0 s \\\hdashline
 & 3 & $10^1$ & \texttt{unsat} &  &  &  & 0.1 s &  & 0.3 s &  & 3 & $10^1$ & \texttt{unsat} & 38 s &  & $>2$ h & 0.9 s &  & $>2$ h \\
 &  & $10^2$ & \texttt{unsat} &  & $\rightarrow$ &  & 0.1 s &  & 0.3 s &  &  & $10^2$ & \texttt{unsat} & 13 m &  & $>2$ h & 16 m &  & $>2$ h \\
 &  & $10^3$ & \texttt{sat} &  &  &  & 0.1 s &  & 0 s &  &  & $10^3$ & \texttt{sat} & 6.8 s &  & 0.3 s & 8.3 s &  & 0.3 s \\\hdashline
 & 4 & $10^1$ & \texttt{unsat} &  &  &  & 0.1 s &  & 3.5 s &  & 4 & $10^1$ & \texttt{unsat} & 2.2 m &  & $>2$ h & 1.1 s &  & $>2$ h \\
 &  & $10^2$ & \texttt{unsat} &  & $\rightarrow$ &  & 0.2 s &  & 3.5 s &  &  & $10^2$ & \texttt{unsat} & 1.1 h &  & $>2$ h & 1 h &  & $>2$ h \\
 &  & $10^3$ & \texttt{sat} &  &  &  & 0.1 s &  & 0 s &  &  & $10^3$ & \texttt{sat} & 13 s &  & 2.1 s & 16 s &  & 2.1 s \\\hline
2 & 2 & $10^1$ & \texttt{unsat} & 0.4 s &  & 0.6 s & 0.2 s &  & 0.9 s & 4 & 2 & $10^1$ & \texttt{unsat} & 3.6 m &  & 33 s & 1.2 s &  & 9.3 m \\
 &  & $10^2$ & \texttt{unsat} & 2.2 s &  & 0.6 s & 2.5 s &  & 0.9 s &  &  & $10^2$ & \texttt{unsat} & 13 m &  & 33 s & 21 m &  & 9.3 m \\
 &  & $10^3$ & \texttt{sat} & 0.4 s &  & 0 s & 0.4 s &  & 0 s &  &  & $10^3$ & \texttt{sat} & 25 s &  & 0.1 s & 39 s &  & 0.1 s \\\hdashline
 & 3 & $10^1$ & \texttt{unsat} & 0.8 s &  & 1.8 m & 0.3 s &  & 3.5 m &  & 3 & $10^1$ & \texttt{unsat} & 31 m &  & $>2$ h & 1.4 s &  & $>2$ h \\
 &  & $10^2$ & \texttt{unsat} & 8.9 s &  & 1.8 m & 9.2 s &  & 3.5 m &  &  & $10^2$ & \texttt{unsat} & $>2$ h &  & $>2$ h & $>2$ h &  & $>2$ h \\
 &  & $10^3$ & \texttt{sat} & 0.7 s &  & 0.1 s & 0.7 s &  & 0.1 s &  &  & $10^3$ & \texttt{sat} & 51 s &  & 0.6 s & 1.4 m &  & 0.6 s \\\hdashline
 & 4 & $10^1$ & \texttt{unsat} & 1.6 s &  & $>2$ h & 0.5 s &  & $>2$ h &  & 4 & $10^1$ & \texttt{unsat} & $>2$ h &  & $>2$ h & 2.1 s &  & $>2$ h \\
 &  & $10^2$ & \texttt{unsat} & 26 s &  & $>2$ h & 29 s &  & $>2$ h &  &  & $10^2$ & \texttt{unsat} & $>2$ h &  & $>2$ h & $>2$ h &  & $>2$ h \\
 &  & $10^3$ & \texttt{sat} & 1.1 s &  & 0.7 s & 1.1 s &  & 0.7 s &  &  & $10^3$ & \texttt{sat} & 2 m &  & 4.3 s & 3.9 m &  & 4.3 s \\\hline
\end{tabular}

    \smallskip

    According to $N_S$, total number of nodes is
    \(2 \mapsto 10, 3 \mapsto 17, 4 \mapsto 26\),
    and more importantly, the number of possible paths for each train is
    \(2 \mapsto 4, 3 \mapsto 27, 4 \mapsto 256\).
}
}

\newcommand{\TextResultsNop}{%
\full{%
Firstly, let's see cases with scenario \var{nop},
that is, with no schedule constraints.
}%
The results are shown in \refTab{exp:nop}.
Since all the results are \Sat{},
it is proved that all trains can finish within the corresponding $J$~steps
(i.e. \refEq{phi:finish} holds).
We additionally show the number of conflicts
which the SAT solver of our tool had to resolve with a~backjump.
Note that all these conflicts are unrelated to schedule constraints,
but are consequences of violations of mutual exclusion conditions instead,
as discussed earlier at \refImg{consecutive}.
}

\full{%
We present the qualitative results (\Sat{} or \Unsat{})
along with the run-times of the tools.
Discussion and interpretation of the results follows in the next paragraph.

\TextResultsNop
}

\full[The ]{%
Concrete results of the scenarios \var{last} and \var{all}
are shown in \refTab{exp:lt}.
\hide{%
$\circ$ is being only~$<$ here,
because $>$ is not supported by \emph{rail\-perf\-check}.
}%
Most importantly, the
}%
results of all the specified case studies\full{ of these scenarios} are as follows:
\begin{itemize}
\item \Unsat{} when \(\var{bnd} \leq 10^2\):
    cases with lower timing upper bound are unsatisfiable,
    that is, it is impossible for the trains to finish within this time bound,
\item \Sat{} when \(\var{bnd} = 10^3\):
    plans with high timing upper bound do exist.
\end{itemize}
\short{%
To give a~basic idea on the run-times of particular experiments
in the case of our approach,
the satisfiable cases do not exceed 4~minutes
and the unsatisfiable cases usually do not exceed 2~hours.
}

\newcommand{\TextResultsLtAbs}{%
The scenarios \var{last_{abs}} and \var{all_{abs}} are included as well,
to~show that in such cases,
our method prunes the searched state space even more, based on the upper bound \var{bnd}.
}%

\hide{%
\TextResultsLtAbs
}

\newcommand{\TableResultsGt}{%
\Table{exp:gt}{0.9}
{Running time of \var{last} and \var{all} scenarios with timing $>\var{bnd}$}
{\CsvTabular{gt}{rrr|l||l|l|l|l}}
}

\newcommand{\TextResultsGt}{%
The rest of selected case studies cover cases with $\circ$ being~$>$,
and is shown in \refTab{exp:gt}.
}

\hide{%
\TableResultsGt

\TextResultsGt
}

\paragraph{Discussion.}
First, we compare the scenarios \var{last} and \var{all}.
In the satisfiable cases, run-times of both scenarios are similar.
In the unsatisfiable cases,
the run-time of scenario \var{all} is generally longer than that of \var{last},
because the trains are unordered
and all their permutations are tried,
\full{which is }a~significant effort with multiple trains.
On the other hand,
to~detect that the relative timing of the last train in scenario \var{last} is unfeasible,
all the preceding trains have to be simulated first, regardless the timing.
Depending on the value of \full{the timing upper bound~}\var{bnd},
it is not certain which part will dominate the run-time---%
simulations of the preceding trains, or of the last train.
\full{%
More on this comparison follows later.
}%
\hide{%
absolute timing experiments \var{\cdot_{abs}}
}

Next, we investigate the behavior of our tool and the tool \emph{rail\-perf\-check}~\cite{rail-art}.
We start with \emph{rail\-perf\-check}.
Recall that it handles mutual exclusion conditions using signal interlocking,
which is efficient for networks that do use signals.
Moreover, they do simulations in lazy offline fashion,
that is, only after a~full propositional assignment was found.
This is especially efficient in the presented satisfiable cases.
However, within the unsatisfiable cases\full{
of the scenarios \var{last} and \var{all}%
},
it is entirely insensitive to the value of \var{bnd},
because only the overall simulation is checked,
independently from whether it satisfies the timing or not.
In this way, early detection of unsatisfiability is not possible,
and all $N_P^{N_S}$ choices of paths are always examined.

In our case,
the value of \var{bnd} has significant impact
on pruning the searched state space\full{ in our case}---%
\refEq{phi:sched:time:rel:<} ensures termination of all search attempts
where it is already obvious that the timing cannot be satisfied.
As a~result,
with growing size of the network and the set of trains,
unsatisfiability is detected more efficiently by our more sophisticated algorithm.
Consequently, if the timing upper bound is low (\(\var{bnd} = 10^1\)),
scenario \var{all} is always faster than scenario \var{last} in our case,
because the unfeasible timing of the train that enters first in the case of scenario \var{all}
can be detected sooner than that in the case of scenario \var{last},
where the train that must satisfy the timing is the last one (as discussed above).
For example, the run-time of our tool in the cases with \(N_T = N_S = 3\),
\(\var{bnd} = 10^1\)
was 1~s and 38~s in the case of scenario \var{all} and \var{last}, respectively,
while with \(\var{bnd} = 10^2\), it was approximately 15~min in both cases.
\full{%
Also, in the case of just one train (\(N_T = 1\)),
our algorithm detects the symmetry of parallel railroads,
and prunes alternative routes.
This is especially significant in the cases
with large infrastructure (\(N_S = 4\))\full{ in \refTab{exp:lt}}.
}

When multiple trains drive consecutively,
our method suffers from a~number of mutual exclusion conflicts
(\refEq{phi:model:mutual}).
For example, in the case of the conflict captured in \refImg{consecutive},
we resolve it by backtracking the whole situation
and seeking another plan where train~$B$ enters later.
The tool \emph{rail\-perf\-check} prevents such a~conflict implicitly within the simulator---%
by stopping train~$B$ at node~1 (if there is a~signal)
until the conflicting section becomes free.
If the signal was not there,
such a~plan of two consecutive trains would not even be considered.

\newcommand{\TextResultsDiscussionNop}{%
Thus, it is not surprising that \emph{rail\-perf\-check} is much faster\full[,
due to the discussion in \refSec{s:exp}%
]{%
in the satisfiable cases,
especially in the scenario \var{nop}%
}.
Our model and/or algorithm should be improved
so that such a~scenario, with no schedule constraints,
is trivial to be solved, like in the case of~\cite{rail-art}.
}

\full{%
\TextResultsDiscussionNop
}

\newcommand{\TextResultsDiscussionLt}{%
When comparing absolute run-times of our tool and \emph{rail\-perf\-check},
we would like to emphasize that the tools solve different problems,
and even the underlying scenarios\full{ \var{last} and \var{all}}
of the corresponding case studies
are not equivalent in cases with multiple trains (\(N_T > 1\)).
The issues stem from the way how mutual exclusivity of trains is handled.
In the case of \emph{rail\-perf\-check},
it is based on signal interlocking,
where whole sections surrounded by signals are being allocated for each train.
Since the presented case studies do not require (nor allow) the trains to stop at stations,
it would be necessary not to put signals in the place of stations.
However, this would also mean that each train always allocates the whole network,
so the next one would have to wait until the train that is ahead exits at the boundary,
resulting in way too long gaps between trains,
and thus in satisfying the timing constraints under different conditions.
To avoid this, we placed signals at stations.
However, this allows situations like when a~train stops in a~station
and waits until another one overtakes it.
While this does not harm the performance in the case of the satisfiable cases,
it has quite significant effect on the run-time performance of the unsatisfiable cases,
because all combinations of particular overtakings of trains are tried,
in addition to all possible routes.
We are not aware of a~way how to avoid these overtakings, though,
even using additional ordering constraints,
which do not seem to be possible to be encoded in \emph{rail\-perf\-check}
in the case of scenario \emph{all},
because they do not allow Boolean combinations of particular ordering constraints.
Also, we did not allow stopping at stations in our case,
because otherwise it would yield an~even more difficult problem (with more nondeterminism)
than in the case of \emph{rail\-perf\-check},
since we cannot set deterministic waiting times at stations,
nor make the trains stop at stations only when it is necessary to avoid a~conflict of trains.

In \refTab{exp:lt}, one can see that at least in the case of a~single train (\(N_T = 1\)),
where both approaches actually yield comparable problems,
our approach perform significantly faster within the unsatisfiable cases.
\short{%
Within all satisfiable cases,
it is \emph{rail\-perf\-check} with the significantly faster run-times, though.
}%
In the rest unsatisfiable cases,
our approach is usually faster,
but one has to take the differences of both approaches into consideration.
Nevertheless, in the case of low timing upper bound (\(\var{bnd} = 10^1\)),
it is expectable that even if \emph{rail\-perf\-check}
did not consider plans with overtakings of the trains,
our approach would still be faster,
because we prune the searched state significantly there.
}

\full{%
\TextResultsDiscussionLt
}

\newcommand{\ParPossibleImprovements}{%
\paragraph{Possible improvements.}
In cases of a~missed deadline by a~train,
a~possibility is to learn that all other trains,
that are \emph{not faster}\footnote{%
For example, train~$T_2$ is surely not faster than train~$T_1$
if \(\xth{A}{T_2} \leq \xth{A}{T_1} \land \xth{B}{T_2} \leq \xth{B}{T_1}
\land \xth{\var{V_{max}}}{T_2} \leq \xth{\var{V_{max}}}{T_1}
\land \xth{L}{T_2} \geq \xth{L}{T_1} \).
}
than the one that participates in the conflict,
would miss the deadline too, if choosing the same route (if it is a~part of their connection),
and if they are constrained by such a~deadline as well.
Such more sophisticated conflict reasoning
would enhance, for example, solving the scenario \var{all} a~lot,
because it would avoid trying all the permutations of the trains, in the unsatisfiable cases.

\hide{%
Also, again in the case of a~missed deadline,
the learning can be extended such that the train would miss the deadline
even if choosing any other path terminating at the same node as the conflicting route,
if driving such a~route would not be faster than in the case of the conflict.
However, it can be challenging to compare driving times of different paths
(considering that the dynamics of trains can be complex) without actually simulating them,
but at least some special cases can be identified,
or some kind of dynamic programming can be used
(e.g. not simulating paths that were already simulated,
or where all parts of the path were already simulated and their driving times can be put together).
Other way is to, at least, only estimate driving times of paths
and use it for purposes of decision heuristic, which can be helpful in the satisfiable cases.
}

\hide{%
In the case of complex formulas where the number of unrollings~$J$ must be high,
\refEq{phi:sched:time:rel} can become large
(as stated in\full[%
{\open[ \refAppend{aa:phi:size}]{~\cite{my-rail-online-art}}}%
]{
\refSec{ss:phi:size}%
}),
and it may be considerable to handle these constraints lazily.
The thing is that it is only a~small part of \refEq{phi:sched:time:rel:<}
that gets actually propagated.
A~way how to handle these constraints is to extend theory propagations
such that the right hand side of the implication in \refEq{phi:sched:time:rel:<}
is learned as soon as the left hand side of the implication---violation of a~timing---%
is true.
}

Finally, the synchronous model is not efficient,
because the granularity of the discretization is too high.
Furthermore, when resolving a~conflict,
the solver backtracks to a~previous state,
causing to also throw away simulations that had nothing to~do with the conflict.
This would not happen in an~asynchronous model.
It should be replaced by an~asynchronous model,
with embedding some timing constraints
also into the systems of ODEs,
to~achieve proper synchronization\footnote{%
This would also allow efficient handling
of timings with the relational operator~$=$.
}.
Note that this does not concern the algorithm,
only the encoding of a~formula.
}

\full{\ParPossibleImprovements}

%%%%%%%%%%%%%%%%%%%%%%%%%%%%%%%%%%%%%%%%%%%%%%%%%%%%%%%%%

\section{Conclusion}\label{s:conclusion}
We presented a~formalization of a~low-level railway scheduling problem,
where the dynamics of trains is described by differential equations,
and where rich timing and ordering constraints are supported.
We analyzed the behavior of our approach compared to an~existing method
on selected case studies, and identified strong and weak aspects of the run-time performance.
We demonstrated that despite the complexity of our model,
the resulting problems can be solved successfully within a~SAT modulo theory framework.
This opens the possibility of applying such techniques to~further application domains
with similar complexity.

%%%%%%%%%%%%%%%%%%%%%%%%%%%%%%%%%%%%%%%%%%%%%%%%%%%%%%%%%

\open{%
\subsection*{Acknowledgements}\label{s:ack}
The work of Stefan Ratschan was supported by
the project GA21-09458S of the Czech Science Foundation GA ČR
and institutional support RVO:67985807.
The work of Tomáš Kolárik was supported by CTU project SGS20/211/OHK3/3T/18.
}

%%%%%%%%%%%%%%%%%%%%%%%%%%%%%%%%%%%%%%%%%%%%%%%%%%%%%%%%%
%%%%%%%%%%%%%%%%%%%%%%%%%%%%%%%%%%%%%%%%%%%%%%%%%%%%%%%%%

\clearpage
\bibliographystyle{splncs04}
\bibliography{main}

\begin{thebibliography}{10}
\providecommand{\url}[1]{\texttt{#1}}
\providecommand{\urlprefix}{URL }
\providecommand{\doi}[1]{https://doi.org/#1}

\bibitem{odeint-art}
Ahnert, K., Mulansky, M.: Odeint~--~solving ordinary differential equations in
  {C}++. {AIP} Conf. Proc. 1389 pp. 1586--1589 (2011),
  \url{https://doi.org/10.1063/1.3637934}

\bibitem{rail-sched-micro-macro}
Albrecht, T.: Railway timetable and traffic. Eurailpress: Hamburg, Germany
  (2008)

\bibitem{multi-agent-art}
Andreychuk, A., Yakovlev, K., Surynek, P., Atzmon, D., Stern, R.: Multi-agent
  pathfinding with continuous time. Artificial Intelligence  \textbf{305},
  103662 (2022), \url{https://doi.org/10.1016/j.artint.2022.103662}

\bibitem{smtlib-reference-art}
Barrett, C., Fontaine, P., Tinelli, C.: The {SMT-LIB} standard, version 2.6
  (2017),
  \url{http://smtlib.cs.uiowa.edu/papers/smt-lib-reference-v2.6-r2017-07-18.pdf}

\bibitem{bmc-art}
Biere, A.: Bounded model checking. In: Biere, A., Heule, M., van Maaren, H.,
  Walsh, T. (eds.) {Handbook of Satisfiability}, chap.~14, pp. 457--481. IOS
  Press (2009), \url{https://doi.org/10.3233/978-1-58603-929-5-457}

\bibitem{continuous-survey}
Bournez, O., Campagnolo, M.L.: A survey on continuous time computations. In:
  Cooper, S., L{\"o}we, B., Sorbi, A. (eds.) New Computational Paradigms, pp.
  383--423. Springer New York (2008),
  \url{https://doi.org/10.1007/978-0-387-68546-5\_17}

\bibitem{float-art}
Brain, M., Tinelli, C., R{\"u}mmer, P., Wahl, T.: An automatable formal
  semantics for {IEEE}-754 floating-point arithmetic. In: 22nd {IEEE} Symposium
  on Computer Arithmetic. pp. 160--167. IEEE (2015),
  \url{https://doi.org/10.1109/ARITH.2015.26}

\bibitem{isat-ode-art}
Eggers, A., Ramdani, N., et~al.: Improving {SAT} modulo {ODE} for hybrid
  systems analysis by combining different enclosure methods. International
  Conference on Software Engineering and Formal Methods ({SEFM})  \textbf{9}
  (2011), \url{https://doi.org/10.1007/978-3-642-24690-6\_13}

\bibitem{minisat}
Eén, N., Sörensson, N.: Mini{SAT} (2008), \url{{http://minisat.se}}

\bibitem{spaceex-art}
Frehse, G., Le~Guernic, C., Donz{\'e}, A., Cotton, S., Ray, R., Lebeltel, O.,
  Ripado, R., Girard, A., Dang, T., Maler, O.: {SpaceEx}: Scalable verification
  of hybrid systems. In: Gopalakrishnan, G., Qadeer, S. (eds.) Computer Aided
  Verification. pp. 379--395. Springer-Verlag, Berlin, Heidelberg (2011),
  \url{https://doi.org/10.1007/978-3-642-22110-1\_30}

\bibitem{dreal-art}
Gao, S., Kong, S., Clarke, E.M.: {dReal}: An {SMT} solver for nonlinear
  theories over the reals. In: Proceedings of the 24th International Conference
  on Automated Deduction. pp. 208--214. CADE'13, Springer-Verlag, Berlin,
  Heidelberg (2013), \url{https://doi.org/10.1007/978-3-642-38574-2\_14}

\bibitem{rail-sched-fixed-timetables-art}
Haehn, R., {\'A}brah{\'a}m, E., Nie{\ss}en, N.: Freight train scheduling in
  railway systems. In: Hermanns, H. (ed.) Measurement, Modelling and Evaluation
  of Computing Systems. pp. 225--241. Springer International Publishing, Cham
  (2020), \url{https://doi.org/10.1007/978-3-030-43024-5\_14}

\bibitem{rail-sched-prob-macro}
Haehn, R., Ábrahám, E., Nießen, N.: Symbolic simulation of railway
  timetables under consideration of stochastic dependencies. In: LNCS. vol.
  12846, pp. 257--275. Springer (2021),
  \url{https://doi.org/10.1007/978-3-030-85172-9\_14}

\bibitem{my-tap-art}
Kol{\'a}rik, T., Ratschan, S.: {SAT} modulo differential equation simulations.
  In: Ahrendt, W., Wehrheim, H. (eds.) Tests and Proofs. LNCS, vol. 12165.
  Springer (2020), \url{https://doi.org/10.1007/978-3-030-50995-8\_5}

\bibitem{unsot-core_lang}
Kolárik, T.: {UN/SOT} core input language specification (2019),
  \url{https://gitlab.com/Tomaqa/unsot/-/blob/master/doc/lang/core.pdf}

\bibitem{unsot}
Kolárik, T.: {UN/SOT} ({UN/SAT} modulo {ODES} {N}ot {SOT}) (2020),
  \url{https://gitlab.com/Tomaqa/unsot}

\bibitem{unsot-preprocess_lang}
Kolárik, T.: {UN/SOT} preprocessing language (2022),
  \url{https://gitlab.com/Tomaqa/unsot/-/blob/master/doc/lang/preprocess.pdf}

\bibitem{rail-art}
Luteberget, B., Claessen, K., Johansen, C.: {SAT} modulo discrete event
  simulation applied to railway design capacity analysis. Formal Methods in
  System Design  \textbf{57},  211--245 (2021),
  \url{https://doi.org/10.1007/s10703-021-00368-2}

\bibitem{double-vertex-art}
Montigel, M.: Formal representation of track topologies by double vertex
  graphs. In: Proceedings of Railcomp 92 held in Washington DC, Computers in
  Railways 3. vol.~2. Computational Mechanics Publications (1992)

\bibitem{smt-dpll_t}
Nieuwenhuis, R., Oliveras, A., Tinelli, C.: Solving {SAT} and {SAT} modulo
  theories: from an abstract {D}avis--{P}utnam--{L}ogemann-{L}oveland procedure
  to {DPLL(T)}. Journal of the ACM (JACM)  \textbf{53}(6),  937--977 (2006),
  \url{https://doi.org/10.1145/1217856.1217859}

\bibitem{rail-mapf-art}
Salerno, M., Fuentetaja, R., Gragera, A., Pozanco, A., Borrajo, D., et~al.:
  Train route planning as a multi-agent path finding problem. In: Conference of
  the Spanish Association for Artificial Intelligence. pp. 237--246. Springer
  (2021), \url{https://doi.org/10.1007/978-3-030-85713-4\_23}

\bibitem{rail-networks-art}
Schlechte, T., Borndörfer, R., Erol, B., Graffagnino, T., Swarat, E.:
  Micro–macro transformation of railway networks. Journal of Rail Transport
  Planning \& Management  \textbf{1}(1),  38--48 (2011),
  \url{https://doi.org/10.1016/j.jrtpm.2011.09.001}

\bibitem{rail-sched-prob}
Schwanh{\"a}u{\ss}er, W.: Die {B}emessung der {P}ufferzeiten im
  {F}ahrplangef{\"u}ge der {E}isenbahn. Ph.D. thesis (1974),
  \url{https://www.via.rwth-aachen.de/downloads/Dissertation\_Schwanhaeusser\_2te\_Auflage\_Text.pdf}

\bibitem{rail-sched-art}
Wei{\ss}, R., Opitz, J., Nachtigall, K.: A novel approach to strategic planning
  of rail freight transport. In: Helber, S., Breitner, M., R{\"o}sch, D.,
  Sch{\"o}n, C., Graf von~der Schulenburg, J.M., Sibbertsen, P., Steinbach, M.,
  Weber, S., Wolter, A. (eds.) Operations Research Proceedings 2012. pp.
  463--468. Springer International Publishing, Cham (2014),
  \url{https://doi.org/10.1007/978-3-319-00795-3\_69}

\end{thebibliography}

\printglossary[type=\acronymtype]

%%%%%%%%%%%%%%%%%%%%%%%%%%%%%%%%%%%%%%%%%%%%%%%%%%%%%%%%%
%%%%%%%%%%%%%%%%%%%%%%%%%%%%%%%%%%%%%%%%%%%%%%%%%%%%%%%%%

\short{\blind{%

\clearpage
\appendix
\ifSpringer
\chapter*{Appendix}
\fi

%%%%%%%%%%%%%%%%%%%%%%%%%%%%%%%%%%%%%%%%%%%%%%%%%%%%%%%%%

\section{Details on Encoding and Formalization}\label{a:phi}
This section provides additional formulas
and other details to the formalization from \refSec{s:phi}.

%%%%%%%%%%%%%%%%%%%%%%%%%%%%

\subsection{Details on Railway Model}\label{aa:phi:model}

%%%%%%%%%%%%%%

\subsubsection{Mode Conditions.}\label{aaa:phi:model:mode}
\TextDynamicsSwitchMode
\TextDynamicsSwitchBrake

%%%%%%%%%%%%%%

\subsubsection{Positional Constraints.}\label{aaa:phi:model:pos}
\TextPosConstraints

\paragraph{More on away conditions.}
\TextAwayConditions

\paragraph{More on transfer constraints.}
\TextTransferBack

%%%%%%%%%%%%%%%%%%%%%%%%%%%%

\SecFormulaSize{aa}

%%%%%%%%%%%%%%%%%%%%%%%%%%%%

\SecInputLanguage{aa}

%%%%%%%%%%%%%%%%%%%%%%%%%%%%%%%%%%%%%%%%%%%%%%%%%%%%%%%%%

\section{Details on \acrl{smt}}\label{a:alg}
This section provides additional, not too formal information to \refSec{s:alg}
for readers that are not experts in the area of algorithms~\cite{smt-dpll_t}
related to \acrf{sat} and to \acrf{smt}.

\ParDpllCdcl

\ParLazyApproach

\ParOnlineApproach

\ParTheoryPropagation

%%%%%%%%%%%%%%%%%%%%%%%%%%%%%%%%%%%%%%%%%%%%%%%%%%%%%%%%%

\section{More on Experimental Part}\label{a:exp}
This section provides extra information and data to \refSec{s:exp},
which can supply a~further insight to the behavior of our approach.

Recall that we provide a~comparison of our approach
with \emph{rail\-perf\-check} tool~\cite{rail-art}.
\TextEquipment

\hide{%
\csvset{myStyle/.append style={
    table head=\hline\csvlinetotablerow\\,
}}}
\Table{exp:lt}{0.9}
    {Running time comparison of \var{last} and \var{all} scenarios\hide{ with timing $<\var{bnd}$}}
    {\hide{\CsvTabular{lt_short}
    {rrr|l|l@{\hspace{-.5ex}}c@{\hspace{-.5ex}}l
           |l@{\hspace{0ex}}c@{\hspace{0ex}}l
    ||rrr|l|l@{\hspace{-.5ex}}c@{\hspace{-.5ex}}l
           |l@{\hspace{0ex}}c@{\hspace{0ex}}l
    }\hide{\CsvTabular{lt}
    {rrr|l||l@{\hspace{-.5ex}}c@{\hspace{-.5ex}}l
           |l@{\hspace{0ex}}c@{\hspace{0ex}}l
           |l|l
    }}}
    \footnotesize

    \smallskip

    According to $N_S$, total number of nodes is
    \(2 \mapsto 10, 3 \mapsto 17, 4 \mapsto 26\),
    and more importantly, the number of possible paths for each train is
    \(2 \mapsto 4, 3 \mapsto 27, 4 \mapsto 256\).
}

\paragraph{Detailed results}
of the case studies presented in \refSec{s:exp},
including run-times,
are shown in \refTab{exp:lt}.

\TextResultsDiscussionLt

\ParPossibleImprovements

\hide{%
\csvset{myStyle/.append style={
    table head=\hline\csvlinetotablerow\\\hline\hline,
}}}
\Table{exp:nop}{0.9}
    {Running time comparison of \var{nop} scenario}
    {\hide{\CsvTabular{nop}{rr|l||ll|l}}
    \footnotesize\begin{tabular}{|rr|l||ll|l|}
\hline
$N_T$ & $N_S$ & Result & Our & conflicts & Their \\\hline\hline
1 & 2 & \texttt{sat} & 0.1 s & 0 & 0 s \\
  & 3 & \texttt{sat} & 0.1 s & 0 & 0 s \\
  & 4 & \texttt{sat} & 0.1 s & 0 & 0 s \\\hline
2 & 2 & \texttt{sat} & 0.4 s & 4 & 0 s \\
  & 3 & \texttt{sat} & 0.6 s & 26 & 0.1 s \\
  & 4 & \texttt{sat} & 1 s & 51 & 0.7 s \\\hline
3 & 2 & \texttt{sat} & 4.3 s & 29 & 0 s \\
  & 3 & \texttt{sat} & 8.3 s & 284 & 0.3 s \\
  & 4 & \texttt{sat} & 17 s & 684 & 2.1 s \\\hline
4 & 2 & \texttt{sat} & 39 s & 106 & 0.1 s \\
  & 3 & \texttt{sat} & 1.5 m & 1559 & 0.6 s \\
  & 4 & \texttt{sat} & 4 m & 5100 & 4.3 s \\\hline
\end{tabular}
}

\paragraph{More experiments.}
\TextSpecNop

\TextResultsNop
\TextResultsDiscussionNop

%%%%%%%%%%%%%%%%%%%%%%%%%%%%%%%%%%%%%%%%%%%%%%%%%%%%%%%%%

}}

%%%%%%%%%%%%%%%%%%%%%%%%%%%%%%%%%%%%%%%%%%%%%%%%%%%%%%%%%
%%%%%%%%%%%%%%%%%%%%%%%%%%%%%%%%%%%%%%%%%%%%%%%%%%%%%%%%%

\end{document}